\documentclass{ab2018}

\usepackage{url}
\usepackage{graphicx}
\usepackage{natbib}
\usepackage{amsmath}
\usepackage{multirow}
\usepackage{longtable}
\usepackage[utf8]{inputenc}
\usepackage[T2A]{fontenc}
\usepackage[russian,english]{babel}

\usepackage{array}
\usepackage{dcolumn}
\usepackage{bm}

\begin{document} 

\title{{\tt MOLLId}: software for automatic identification of spectral molecular lines in the sub-millimeter and millimeter bands and its application to the spectra of protostars from the region RCW 120}

\author{A.~A.~Farafontova$^{1}$, M.~S.~Kirsanova$^{1}$, S.~V.~Salii$^{2}$}

\institute{$^{1}$Institute of Astronomy, Russian Academy of Sciences, Moscow, 119017 Russia
\\
$^{2}$Ural Federal University named after the first President of Russia B. N. Yeltsin, Yekaterinburg, 620062 Russia}

\titlerunning{MOLLId: software for automatic identification of spectral molecular lines}

\authorrunning{Farafontova}

\offprints{Anastasiia Farafontova  \email{farafontovanst@gmail.com} }

\abstract{
In this work, we present the program {\tt MOLLId} (MOLecular Line Identification) for automated molecular lines approximation with gaussian profile. Molecular identification was performed using multi-level comparison of the lines' center frequencies and rest frequencies from the spectroscopic database. The program was tested using identification of the molecular lines in observational spectra of young stellar objects RCW 120 YSO S1 and RCW 120 YSO S2, located near the border of the RCW 120 PDR. In the spectra of the RCW 120 YSO S1 source, 100 lines of 41 molecules were identified over the level of 4-6$\sigma$. In the spectra of the RCW 120 YSO S2 source, 407 lines of 79 molecules were identified over the level 3-5$\sigma$. Using Intel Core i7-12700K CPU, identification time is equal to $\sim$6 and 8 minutes per spectral range for the YSOs S1 and S2, respectively. From the analysis of CH$_{3}$OH, CH$_{3}$CN, CH$_{3}$CCH molecules identified in RCW\,120~YSO~S2 we found a two-component structure and estimated the physical parameters in the LTE approximation for each of the components.

}

\maketitle

\section{Introduction}
\label{intro}
Emission lines of small and complex molecules serve as a key instrument in the estimation of chemical composition and physical parameters of star-forming regions. Observations in the submillimeter and millimeter range give high-resolution spectra with a “forest” of molecular lines in them. The analysis of spectra requires a rigorous search through spectroscopic databases (e.g., CDMS~\citep{cdms}, JPL~\citep{PICKETT1998883}. It includes the approximation of line profiles with the Gaussian function and the estimation of the most probable molecules for every spectral line. Different methods are used for the line identification. Manual search (e.g.~\citet{2014ApJ...786...38H} with {\tt SPLATALOGUE}\footnote{https://splatalogue.online/\#/home}, \citet{2019A&A...631A.142G}) is the most laborious method that is used for visual identification of the lines in spectra. 

In order to simplify the processing of spectral data, different programs and methods are developed. These programs fit lines with one or multiple Gaussian components. One approach can be seen in the work~\citep{2000AA...364...83H}. The main method of this work relies on the continuous addition of Gaussian profiles into the line profile until a certain noise level is reached. On the final stage of the approximation, the number of Gaussian components is reduced by excluding statistically unreasonable components or by uniting spatially related components. Alternative methods use the Autonomous Gaussian Decomposition (AGD) algorithm. These methods are {\tt GAUSSPY}~\citep{Lindner_2015}, {\tt GAUSSPY+}~\citep{Riener}, {\tt SCOUSEPy}~\citep{2016MNRAS.457.2675H}. Generally, this method distinguishes potential Gaussian components in the line profile by calculation of the first and second derivatives of the observed spectra. Another program, e.g., {\tt SPIF}~\citep{juvela} allows the user to approximate line profiles with one or multiple Gaussian profiles. The best fit of the observed profile is found by calculation of $\chi^{2}$. 

Except for the software mentioned above, another type of programs for the line identification and profile approximation is developed. These programs are based on the estimation of the best model spectra solution using local thermodynamical equilibrium (LTE) approximation. Namely, {\tt XCLASS}~\citep{moller}, {\tt PySpecKit}~\citep{2011ascl.soft09001G}, {\tt MADCUBA}~\citep{martin}, {\tt CASSIS}~\citep{2015sf2a.conf..313V} are used. The distinctive feature of the programs {\tt MADCUBA} and {\tt CASSIS} is the presence of a graphical shell and manual selection of the molecules for the LTE approximation. Calculation of the LTE spectra and identification is more accurate than manual identification. However, it requires setting some pre-known conditions, e.g., chemical composition. This approach follows the assumption that the emission of all molecules occurs under the same physical conditions.

Objects such as hot cores are related to the earliest stages of star-formation. During the hot core stage, the protostar is embedded in the surrounding gas, and the temperature is high enough for the evaporation of the molecules from the surfaces of dust grains and enough for the formation of more complex molecular species in the gas phase. The hot core spectrum consists of many emission lines of the small and complex molecules, for example, SO, CS, SO$_{2}$, CH$_{3}$CN, CH$_{2}$(OH)CHO, HCOOCH$_{3}$, CH$_{3}$OCH$_{3}$ and others. 

In this work, we use emission spectra of the two young stellar objects (YSOs) in the millimeter range. These YSOs are located near the border of the RCW\,120 PDR. RCW\,120 is an actively studied region during the last decade. It's located at the distance of 1.34~kpc from the Sun~\citep{2009A&A...496..177D}. Near the border of the PDR many YSOs were found towards the clumps observed in the millimeter range, see~\citet{2007A&A...472..835Z}. Their evolutionary classes were established as 0, I and II. Further works by \citet{figueira} and \citet{2018AA...611A..98B} analyzed IR-sources and estimated masses of the YSOs. It was shown that the source S2 is the most massive among all sources. The mass of the source S1 is several times less. Hereinafter, we will refer to these sources as RCW\,120~YSO~S2 and RCW\,120~YSO~S1, following the notation used in the paper by~\citet{2024AstBu..79..235P}.

\section{Observations}

\begin{figure*}
    \centering
    \includegraphics[width=1.0\linewidth]{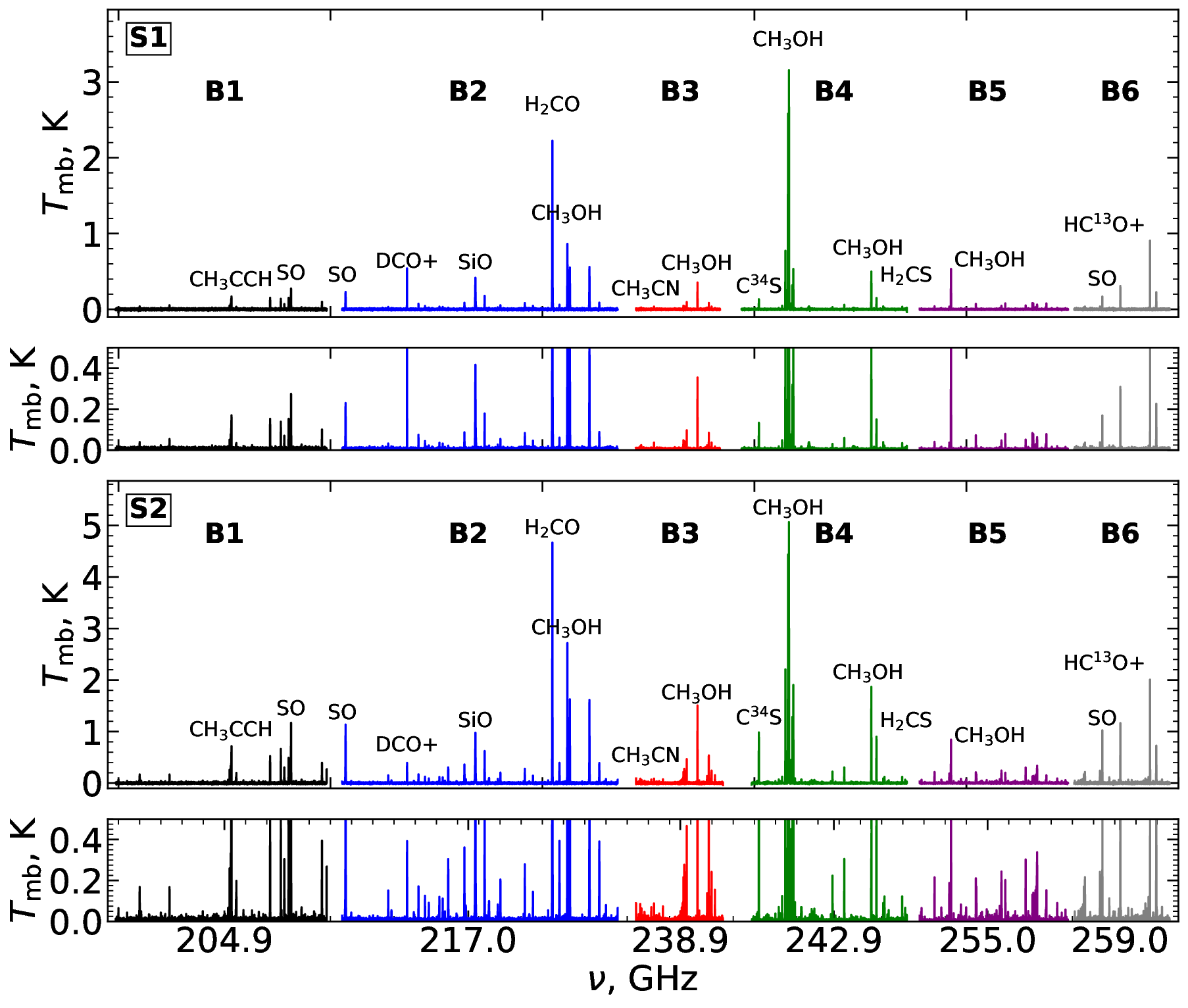}
    \caption{Spectra of RCW\,120~YSO~S1 (two top panels) and RCW\,120~YSO~S2 (two bottom panels). Data corresponding to spectral bands B1 $-$ B6 from Table~\ref{tab:obs_bands} are color-coded. The most intense lines are labeled with the names of the identified molecules}
    \label{fig:observed}
\end{figure*}

Observations were performed using the APEX telescope~\citep{2006AA...454L..13G} in Chile during the periods May 19–21, June 4–5 and 24, August 22–23 and 30, 2022, under program E-0109.C-0623A-2022 (PI: Kirsanova M.~S.), ESO observation time. The main goal of this observation program was to obtain high-quality spectra with a noise level of less than 10 mK at a spectral resolution of at least 0.3 km s$^{-1}$. The nFLASH230 receiver~\citep{2018AA...611A..98B} was used for observations in six frequency bands listed in Table~\ref{tab:obs_bands}. Observational spectra are shown on the figure~\ref{fig:observed}. As a result of signal accumulation over a period of approximately 3-4 hours in each of the bands, the required noise level was achieved.

The data were calibrated to the antenna temperature scale in real time using the standard {\it apexOnlineCalibrator} package. We then converted the spectra to the main beam temperature scale by dividing the spectra by coefficients $\eta_{\rm mb} = 0.81$ and 0.73 for May/June and August, respectively\footnote{https://www.apex-telescope.org/telescope/efficiency/index.php}. The spatial resolution of the observations as a function of frequency was (21-34) arcsec, which corresponds to (0.13-0.22) pc for the distance RCW 120.

The two observed positions with coordinates $\alpha=17\rm ^h$12$\rm ^m$10.960$\rm ^s$, $\delta=-38^\circ$~31\arcmin59.7\arcsec{} and $\alpha=17\rm ^h$12$\rm ^m$08.700$\rm ^s$, $\delta=-38^\circ$~30\arcmin47.4\arcsec{} correspond to the protostars RCW\,120~S1 and RCW\,120~S2. The observations were performed in ON and OFF mode with the OFF position at $\alpha=17\rm ^h$12$\rm ^m$08.000$\rm ^s$, $\delta=-38^\circ$~36\arcmin03.00\arcsec. The precipitated water vapor (PWV) level was in the range $0.3-0.9$ mm, and the noise temperature was $T_{\rm sys} \approx 100-150$ K.

\begin{table}[]
    \centering
    \begin{tabular}{ccc}
    \hline
     Range & Frequencies & Noise level S1/S2\\
               & (GHz)   & (mK) \\
     \hline
     B1        & 202.951 $-$ 206.831 &   6/6\\ 
     B2        & 215.170 $-$ 219.169 &   5/7\\ 
     B3        & 236.921 $-$ 240.920 &   4/6\\ 
     B4        & 240.831 $-$ 244.830 &   3/4\\ 
     B5        & 253.169 $-$ 257.167 &   3/4\\ 
     B6        & 257.081 $-$ 261.078 &   4/7\\ 
     \hline
    \end{tabular}
    \caption{Selected frequency ranges. Noise levels are given for spectra smoothed to a spectral resolution of 0.3 km s$^{-1}$}
    \label{tab:obs_bands}
\end{table}

\section{Methods}
The {\tt MOLLId} software package consists of two stages: spectral line profile approximation and molecule identification. Line profile approximation is performed using a single-component Gaussian function. For each line that stands out at a given noise level, the following parameters are determined using an optimization method: central frequency, line width, and line intensity. The central frequencies obtained in the first stage are used in the second stage for comparison with frequencies from spectroscopic databases.

Before applying the software package to the observation data, baseline subtraction was performed using the {\tt arpls} function from the {\tt pybaselines} package \citep{erb_2024}. This function is based on the asymmetric weighted least squares smoothing approach.

\subsection{Approximation of line profiles}

Spectral line profiles are usually studied using simplified models. The most common approach is to approximate the profile using a Gaussian function. The use of a Gaussian function is based on the hypothesis of a Maxwellian distribution of velocities in a gas, where thermal motion predominates. In a broader context, the central limit theorem states that with a sufficient number of random movements within the directional diagram, caused, for example, by turbulence, the observed spectral lines will approach a Gaussian shape.

In this work, the \texttt{python} programming language was used to implement the process of sequential approximation of spectral line profiles using the Gaussian function. Each line is approximated by a single Gaussian profile with three independent parameters: central frequency ($\rm x_0$), half-width at half maximum ($\sigma^{‘}$), and line intensity maximum ($\frac{1}{\sigma^{’}\sqrt{2\pi}}$) in the form:
\begin{equation}
    f(x_i)=\frac{1}{\sigma^{'}\sqrt{2\pi}}\exp{\left[\frac{(x_i-x_0)^{2}}{2\sigma^{'2}}\right]}.
\end{equation}

As a first approximation, the program uses a single Gaussian profile. Consequently, it does not take into account the case of line overlapping, when more than one source with different velocities can be on the line of sight. In addition, possible line wings are not taken into account. The cases listed above require a more detailed approach than we have currently implemented in {\tt MOLLId}.

The approximation process starts with the line that is the most intensive in the chosen range. This line is found using the \texttt{numpy.argmax} function. Then the line is approximated by a Gaussian profile and subtracted from the spectrum. In the modified spectrum, the most intense line is selected again and approximated by a Gaussian function. The process stops when no lines above the 3$\sigma$ level are distinguished in the spectrum, where $\sigma$ is the noise level in the broadband spectrum. Random outliers are filtered out by calculating the coefficient of determination (R$^{2}$), which is calculated for each approximation:
\begin{equation}
R^2 = 1 - \frac{\sum_{i=1}^{n} e_i^2}{\sum_{i=1}^{n} (y_i - \bar{y})^2},
\end{equation}

where e$_{i}$ is the vector of the residuals, y$_{i}$ is the observed data, $\bar{\rm y}$ is the mean of the observed data. For a random outlier, R$^{2} < 0.1$ is specified.

The parameters for each line are evaluated using the \texttt{least\_squares} optimization function with the \texttt{trf} method from \texttt{scipy.optimize}. The function minimizes the difference $\sum_{i=1}^{n} [y_i - f(x_i)]^2$, where x$_{i}$ $-$ frequency values at point ${i}$, y$_{i}$ $-$ line intensity values from observations, ${n}$ $-$ total number of data points in the range around the line center, $f(x_i)$ $-$ Gaussian function. The values y$_{i}$ and x$_{i}$ are loaded from a text file with the spectrum. The initial value of the line width (FWHM) and noise level (3$\sigma$) are given by user to the input. The sum of the square deviations of the frequency alters after every algorithm iteration. It occurs due to the line substraction from the spectrum after the optimal solution is found.

An adaptive approach is used to estimate the line parameters, namely, the width of the first line found is taken into account to improve the initial approximation for the remaining lines. The uncertainties of the obtained Gaussian parameters are calculated using the Jacobian of the optimization function ($J$), which is a matrix of partial derivatives:
\begin{equation}
J_{ij} = \frac{\partial f(x_i, \vec{p})}{\partial p_j},
\end{equation}
where $\vec{p}$ is the vector of parameters $(A, \mu, \sigma)$ at each i-th step. The standard deviations of the parameters ($\sigma_i$) are calculated using the expression:
\begin{equation}
\sigma_i = \sqrt{\text{diag}((J_{ij}^T J_{ij})^{-1} \cdot \frac{r_i^T \cdot r_i}{n - p})},
\end{equation}
where $n$ is the number of observation data points in the range around the center of the line. The value of $n$ is calculated as an integer value from the result of dividing the line width in GHz by the spectral resolution. The line width is specified at the program input and converted from km s$^{-1}$ to GHz. The value $p$ is the number of model parameters, $r_i$ is the difference between the observation data and the Gaussian approximation for a given solution, and $\text{diag}$ means extracting the diagonal elements of the matrix.

\subsection{Identification of molecules}
The main method used in the second part of the software package is the frequency matching algorithm. The algorithm was implemented taking into account several criteria. The initial selection of molecules is based on frequency, upper level energy ($E_{\rm u}$), and the logarithm of the Einstein coefficient of spontaneous decay ($\log A_{ij}$). We started with the selection levels with the lowest value of $E_{\rm u}$ and the highest value of $\log A_{ij}$, and then moved on to the next level of values if the line could not be identified. Totally, four line selection levels are implemented in the program. The values of $E_{\rm u}$ and $\log A_{ij}$ for each selection level are given in Table~\ref{tab1}. First, the search for molecules is carried out in a narrow spectral range ($\Delta \nu$ $\pm 0.0005$) from the central frequency of the line and for low energies of the upper level ($E_{\rm u}$ $\leq 200$). Then the search boundaries are expanded if no molecules have been selected for a given frequency at this stage. If several candidate molecules have been selected for the same line, the candidate with the minimum frequency deviation from the center, the smallest value of $E_{\rm u}$ and the largest value of $\log A_{ij}$ is selected. The selection is made taking into account the LSR velocity of the object and in accordance with the list of molecules identified in the interstellar medium\footnote{https://cdms.astro.uni-koeln.de/classic/molecules}, as well as taking into account the isotopologues and isotopomers of these molecules.

\begin{table}[!ht]
\caption{Selection criteria used to identify molecules in the spectrum, based on the deviation of the frequency from the spectroscopic database from the central frequency ($\Delta \nu$), the upper level energy ($E_{\rm u}$), and the logarithm of the Einstein coefficient of spontaneous decay ($\log A_{ij}$)}\label{tab1}
\begin{center}
\scriptsize
\begin{tabular}{cccc}
\hline
Level & $\Delta \nu$ (GHz) & $E_{\rm u}$ (K) & $\log A_{ij}$ \\\hline

1 & $\pm 0.0005$ & $\leq 200$ & $\geq 1 \times 10^{-7}$ \\

2 & $\pm 0.001$ & $\leq 400$ & $\geq 1 \times 10^{-8}$ \\

3 & $\pm 0.002$ & $\leq 800$ & $\geq 1 \times 10^{-9}$ \\
4 & $\pm 0.003$ & $\leq 1000$ & $\geq 1 \times 10^{-10}$ \\\hline

\end{tabular}
\end{center}
\end{table}

\subsection{Estimates of physical parameters in the LTE approximation}

To estimate the excitation temperature ($T_{\rm ex}$) and column density ($N$) of molecules in the LTE approximation, we use the rotational diagram method. In this method, the column density of a molecule at the upper level of an optically thin line ($N_{\rm u}$) is determined by the expression:
\begin{equation}\label{roteq1}
    N_\text{u} = \frac{8\pi k \nu^2 \int T_\text{mb} dV }{h c^3 A_{\rm ul}} \quad (\text{cm}^{-2}), \quad
\end{equation}
where $\int T_\text{mb} $dV  is the integral intensity, A$_{ul}$ is the Einstein coefficient of spontaneous decay from the upper to the lower level, $k$ is Boltzmann's constant, $h$ is Planck's constant, $\nu$ is the line frequency in MHz, and $c$ is the speed of light.

Assuming that the excitation temperature of all levels of the molecule is the same and equal to a certain temperature $T_{\rm ex}$, the following expression can be obtained from Boltzmann's formula:
\begin{equation}\label{roteq2}
   \ln\left(\frac{N_{\rm u}}{g_{\rm u}}\right)=\ln\left(\frac{N}{Q_(T_{\rm ex})}\right)-\frac{E_{\rm u}}{kT_{\rm ex}}
\end{equation}
where $E_{\rm u}$ is the upper level energy, $g_{\rm u}$ is the statistical weight of the upper level, and $Q (T_{\rm ex}$) is the statistical sum at excitation temperature T$_\text{ex}$, summed over all levels. From this equality, the expression for N corresponds to: 
\begin{equation}
    N = \frac{N_\text{u}}{g_\text{u}} Q \exp\frac{E_\text{u}}{kT_\text{ex}} \quad (\text{cm}^{-2}). \quad
\end{equation}

To determine the column density, the statistical sum is calculated based on interpolation of data from the spectroscopic database using the approximation:
\begin{equation}
    Q = aT^{b}. \quad
\end{equation}

If there are several emission lines for a given molecule in the observations, the ratio from expression (\ref{roteq1}) is calculated for each line. In this case, based on expression (\ref{roteq2}), the linear function $y=ax+b$ approximates the dependence $\ln\frac{N_{\rm u}}{g_{\rm u}}$ on $E_{u}/k$, whose slope is inversely proportional to $-T_{\rm ex}$, and the intercept is equal to $ \ln\frac{N}{Q}$.

\section{Results}

\subsection{Approximations obtained for line profiles}

\begin{figure}[!ht]
    \centering
    \includegraphics[width=0.95\linewidth]{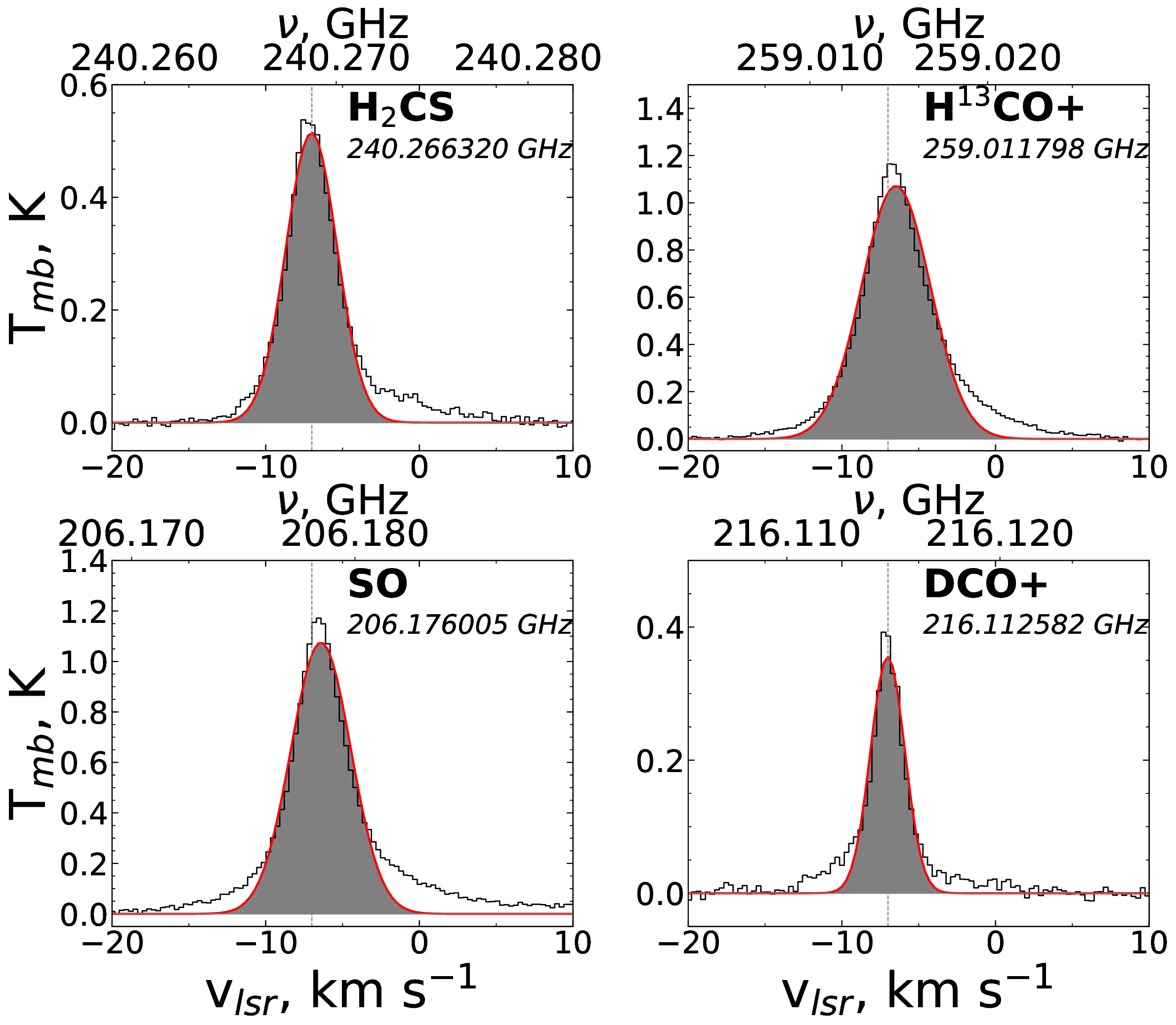}
    
    \caption{Examples of line profile approximations for the RCW\,120~YSO~S2 source. The Gaussian function is shown in red in the figure. The corresponding area under the Gaussian profile is shown in gray}
    \label{fig:profile}
\end{figure}

As a result of applying the {\tt MOLLId} program to the observational spectra of RCW\,120~YSO~S1 and RCW\,120~YSO~S2, approximations of lines with an intensity of 5-7$\sigma$ were obtained. The line parameters for RCW\,120~YSO~S1 and RCW\,120~YSO~S2 are given in Table~\ref{tab:molecular_parameters_core1} and~\ref{tab:molecular_parameters_core2}, respectively. Figure~\ref{fig:profile} shows examples of the obtained approximations of profiles for several molecules. It can be seen that the method has limitations, namely, we do not approximate the wings of lines, which can be prominent for the lines of some molecules. However, this limitation does not affect the identification of lines and the determination of the central frequency and line width. In the tables with the approximation results, it can be seen that the line widths are different. The difference is especially noticeable when comparing lines with low and high (greater than 200 K) $E_{\rm u}$. We discuss the observed difference in the obtained line profile widths and the dependence on the $E_{\rm u}$ in section~\ref{sect:lines}. 

\subsection{Identified molecules}
All molecules identified using the {\tt MOLLId} program in RCW\,120~YSO~S1 and RCW\,120~YSO~S2 are listed in Table~\ref{tab:all_species}. On an Intel Core i7-12700K CPU, the identification time per spectral range was $\sim$ 6 minutes for the RCW\,120~YSO~S1 source and $\sim$ 8 minutes for the RCW\,120~YSO~S2 source. The number of lines per 1 GHz, for example, for the B1 spectral range for RCW\,120~YSO~S1 is $\sim$ 5 lines per 1 GHz, and for RCW\,120~YSO~S2 $-$ 16 lines per 1 GHz.

In the spectra of RCW\,120~YSO~S2, 407 lines of 79 molecules were identified (see Table~\ref{tab:molecular_parameters_core2}). The main source of spectroscopic data ($\approx$80\%) of the lines is the JPL database. Among all frequency ranges, methanol (CH$_{3}$OH) lines are detected most frequently. The number of CH$_{3}$OH lines is 55, corresponding to $\sim$14\% of the total number of lines. Methyl formate (CH$_{3}$OCHO) lines account for 10\% (42 lines) of the total number and are the second most common, CH$_{3}$OCH$_{3}$ accounts for 6\% (23 lines), CH$_{3}$CHO accounts for 5\% (19 lines) of the total number of identified lines. The number of identified methylacetylene (CH$_{3}$CCH) lines is 4\% (18 lines), H$_{2}$CS, CH$_{3}$CN, and sulfur dioxide (SO$_{2}$) account for 4.2\% (17 lines), 3.7\% (15 lines) and 3.4\% (14 lines), respectively. The isotopologue $^{13}$CH$_{3}$OH accounts for 3.4\% of all identified molecules, while SO$^{18}$O accounts for 2.7\% (11 lines). Deuterated species CH$_{2}$DCN account for 1.5\% (6 lines), HDCO 1.0\% (4 lines), CH$_{2}$DCHO 0.7\% (3 lines), and HDO 0.5\% (2 lines) of the total number of molecules. 
\begin{figure}[!ht]
    \centering
    \includegraphics[width=0.95\linewidth]{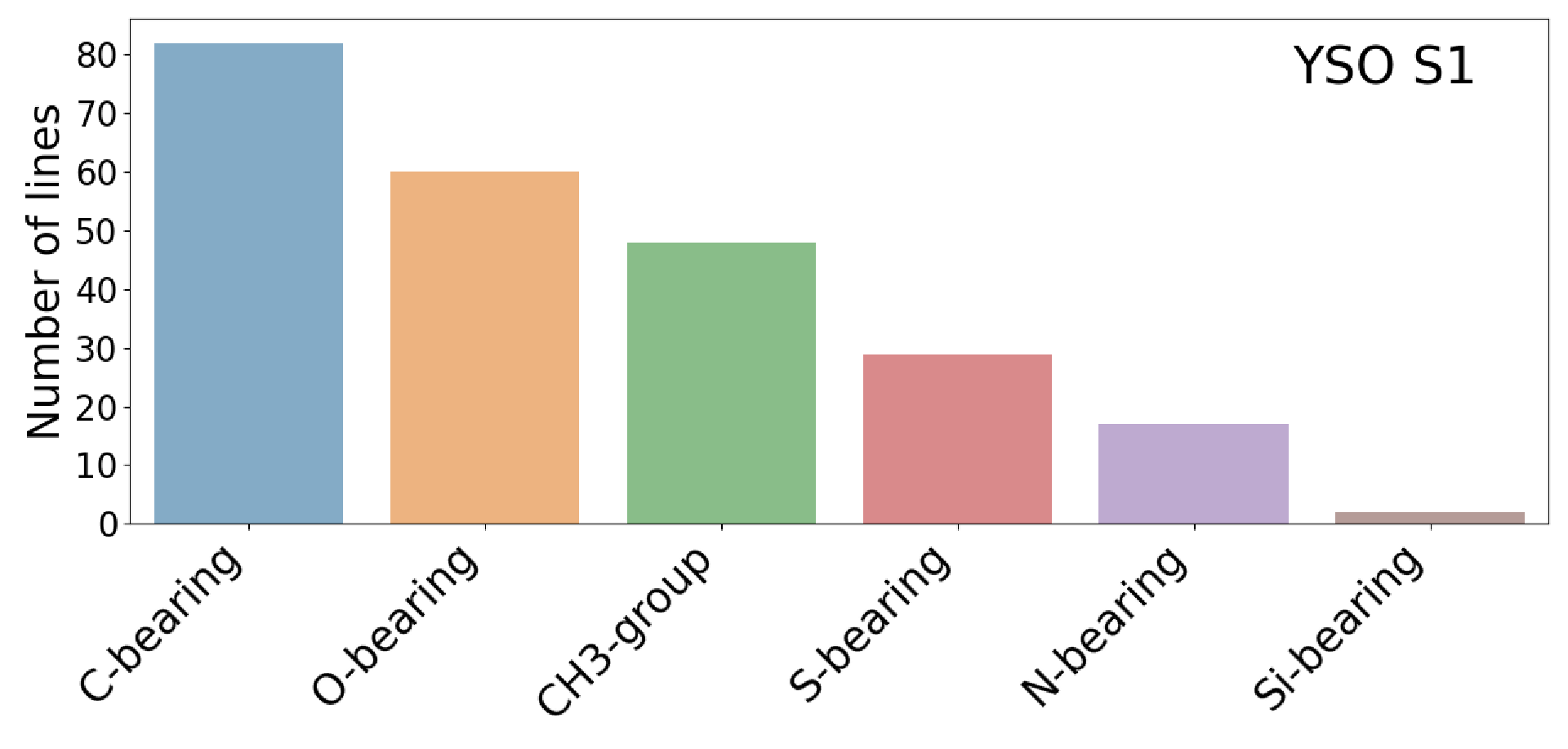}
     \includegraphics[width=0.95\linewidth]{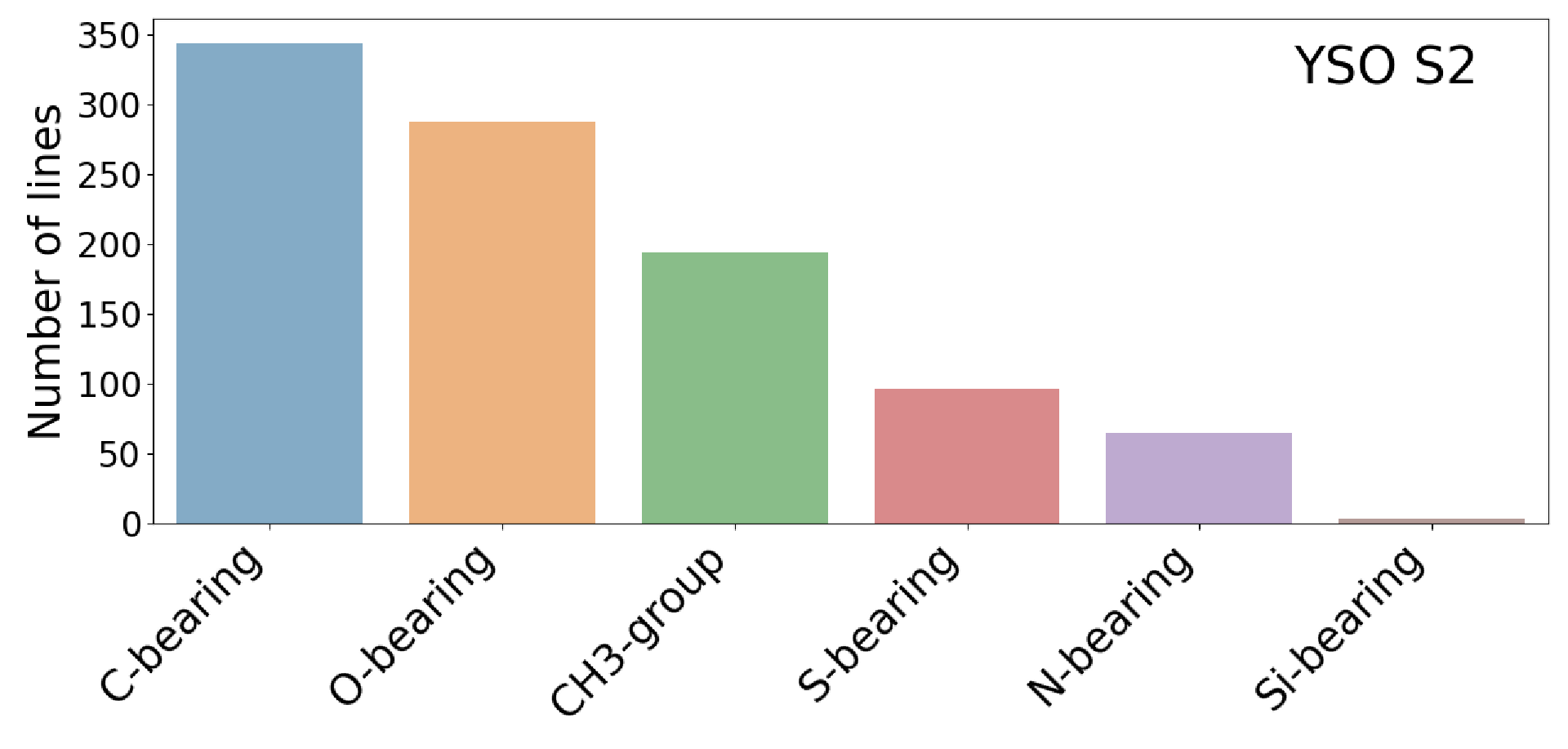}
    
    \caption{Classification by groups for RCW\,120~YSO~S1 (top panel) and RCW\,120~YSO~S2 (bottom panel)}
    \label{fig:groups}
\end{figure}
Classification by groups (see Fig.~\ref{fig:groups}) reveals a large number of carbon-containing molecules, accounting for 84.5\% of the total number of lines. Oxygen-containing species account for 70.8\%, and the CH$_{3}$ group is present in 47.7\% of the total number of lines found. Sulfur-containing species H$_{2}$CS, SO$_{2}$, OCS, HSCN, and their isotopologues account for 23.8\%. Nitrogen-containing species account for 16\%. Silicon monoxide (SiO) and its isotopologues $^{29}$SiO, $^{30}$SiO account for 1\% or 4 lines. 

In the spectra of RCW\,120~YSO~S1, 100 lines of 41 molecules were identified using the developed algorithm (see Table~\ref{tab:molecular_parameters_core1}). Among all molecular lines, the most frequent is CH$_{3}$OH, with 15 identified lines. This number is 3.6 times less than the number of identified CH$_{3}$OH lines in RCW\,120~YSO~S2. The second most frequent molecular line is CH$_{3}$CCH, with 13 lines identified. In the observational spectra, 7 lines of the CH$_{3}$CN molecule, 6 lines of the H$_{2}$CS molecule, and 5 lines of the CH$_{3}$CHO molecule were identified. There are 4 identified lines of SO$_{2}$ and OCS. Number of identified lines of SO, H$_{2}$CO, and c-C$_{3}$H$_{2}$ is 3 for each molecule.

In the final classification by groups (see Fig.~\ref{fig:groups}), 82\% of carbon-containing species, 60\% of oxygen-containing species, 29\% of sulfur-containing species, and 17\% of nitrogen-containing species were identified among the all identified lines. The CH$_{3}$ group accounts for 48\% of all identified lines. The smallest number of lines, 2\%, are molecules containing silicon.  

Quality of molecule line identification depends on the sensitivity threshold used. Thus, for the spectra of RCW\,120~YSO~S2 at a resolution of 0.3~km s$^{-1}$, all lines with amplitudes starting from the $5-7\sigma$ level were identified. Noise levels are specified in Table~\ref{tab:obs_bands} for each range. The spectra contain a large number of spectral components, therefore in the $(3-5)\sigma$ range it is impossible to unambiguously identify overlapping lines of some molecules whose frequencies are too close. For example, there are two lines near the frequency 258726 MHz. The CH$_{3}$OH line with a frequency of 258726.17 MHz and energy 166~K overlaps with the CH$_{3}$OCH$_{3}$ line with a frequency of 258726.56 MHz and energy 88~K. In such a case, we believe that only modeling the spectra for the given parameters of interstellar gas allows us to make a correct choice in favor of one of the two lines or to assume the presence of a blend of two lines. 

\begin{table*}
\centering
\caption{Identified molecules in RCW\,120~YSO~S1 and RCW\,120~YSO~S2 }
\renewcommand{\arraystretch}{0.95}
\begin{tabular}{llllllllll}
\hline
\textbf{2} & \textbf{3} & \textbf{4} & \textbf{5} & \textbf{6} & \textbf{7} & \textbf{8} & \textbf{9} \\
\textbf{atomic} & \textbf{atomic} & \textbf{atomic} & \textbf{atomic} & \textbf{atomic} & \textbf{atomic} & \textbf{atomic} & \textbf{atomic} \\
\hline
\multicolumn{4}{c}{} & \textbf{RCW\,120~YSO~S2} & & & \\
\hline

SO & SO$_2$ & HNCO & H$_2$CCO& CH$_3$OH & CH$_3$CHO & CH$_3$COOH &CH$_{3}$OCH$_{3}$  \\
NS & HDO & H$_2$CS  & H$_2$NCO$^+$ & CH$_3$CN & CH$_3$CCH & CH$_3$OC$^{13}$HO&C$_{2}$H$_{5}$CN \\
SiO & H$_2$S & H$_2$CO &c-C$_3$H$_2$ & C$^{13}$H$_3$OH&HOCH$_2$CN  &  CH$_3$C$_3$N& \\
SO$^+$ & HCO & HDCO & HCOOH & CH$_2$DCN & CH$_2$DCHO&CH$_3$OCHO  & \\
$^{33}$SO & CCH & HSCN & HCC$^{13}$CN & CHD$_2$OH & C$_2$H$_3$CN &  & \\
$^{34}$SO & DCO$^+$ & HDCS & l-CCC$^{13}$H$_2$ &CH$_3$CN, $v$$_{8}$=1 & CHD$_2$CHO& & \\
C$^{34}$S &HSO & H$_2$C$^{34}$S  & HCCC$^{13}$N & l-HC$_4$N & & &  \\
$^{29}$SiO&CCD & H$_2$C$^{33}$S  & CCC$^{13}$CH &H$_2$CNCN & & &  \\
$^{30}$SiO&SO$^{17}$O&c-C$_3$H & C$_4$H & & &  &  \\
& SO$^{18}$O & HCCS & HCCC$^{15}$N& & & & \\
 & $^{33}$SO$_2$ & H$_2$$^{13}$CO &c-HCOOH  &  & & & \\
 & $^{34}$SO$_2$ & HNSO &c-HC$^{13}$CCH &  & & & \\
 & O$^{13}$CS & &  H$_2$CNH &  & & & \\
 & HC$^{18}$O$^{+}$ & &  &  & & & \\
 & H$^{13}$CO$^+$ & & & & & & \\
 & OC$^{34}$S & & &  & & & \\
 & HCN &  & & & & & \\
  & HC$^{13}$N & & & & & & \\
  & HC$^{15}$N & & & & & & \\
  & OCS & & & & & & \\
\hline
\multicolumn{4}{c}{} & \textbf{RCW\,120~YSO~S1} & & & \\
\hline
SO & SO$_2$ & HNCO & H$_2$CCO& CH$_3$OH & CH$_3$CHO & CH$_3$COOH & \\
NS & H$_2$S & H$_2$CS  &c-C$_3$H$_2$  & CH$_3$CN & CH$_3$CCH & CH$_3$O$^{13}$CHO & \\
SiO & HCO & H$_2$CO &H$_2$CNH & CH$_2$DCN&  &  & \\
$^{33}$SO & DCO$^+$ & HDCO &  &  &  &  & \\
 $^{34}$SO& SO$^{17}$O & c-C$_3$H &  &  &  &  & \\
C$^{34}$S & $^{33}$SO$_2$ & H$_2$$^{13}$CO &  & & & & \\
 & HC$^{18}$O$^{+}$& HN$^{13}$CO&   &  & & & \\
& H$^{13}$CO$^+$ & &  & & &  &  \\
& HC$^{13}$N  &  &  &  & & & \\
 & HC$^{15}$N &  & &  & & & \\
 &OCS  & &   &  & & & \\
\hline
\end{tabular}
\label{tab:all_species}
\end{table*}

\subsection{High excitation lines in RCW\,120~YSO~S2}\label{sect:lines}

\begin{figure}[!ht]
    \centering
    \includegraphics[width=0.95\linewidth]{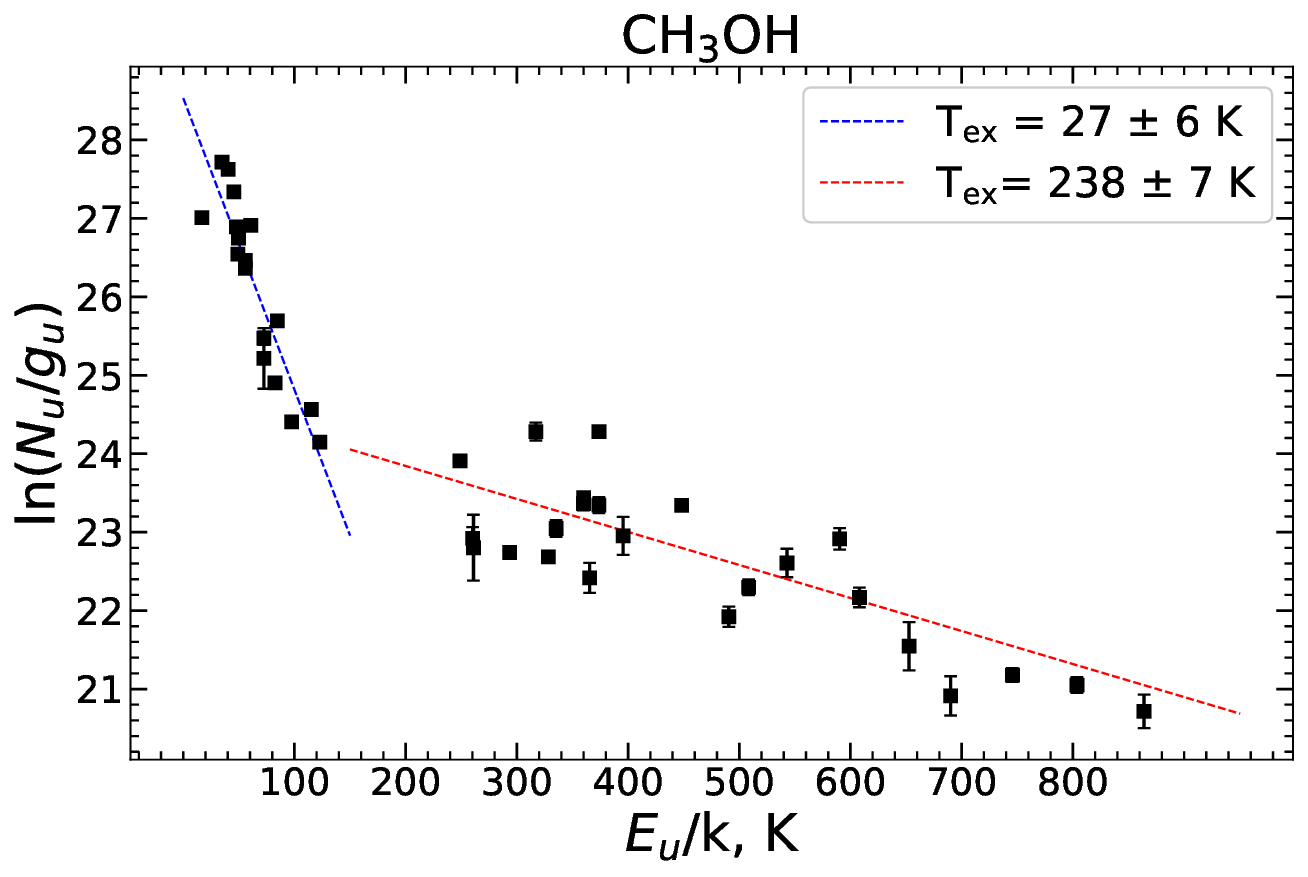}
    \includegraphics[width=1.0\linewidth]{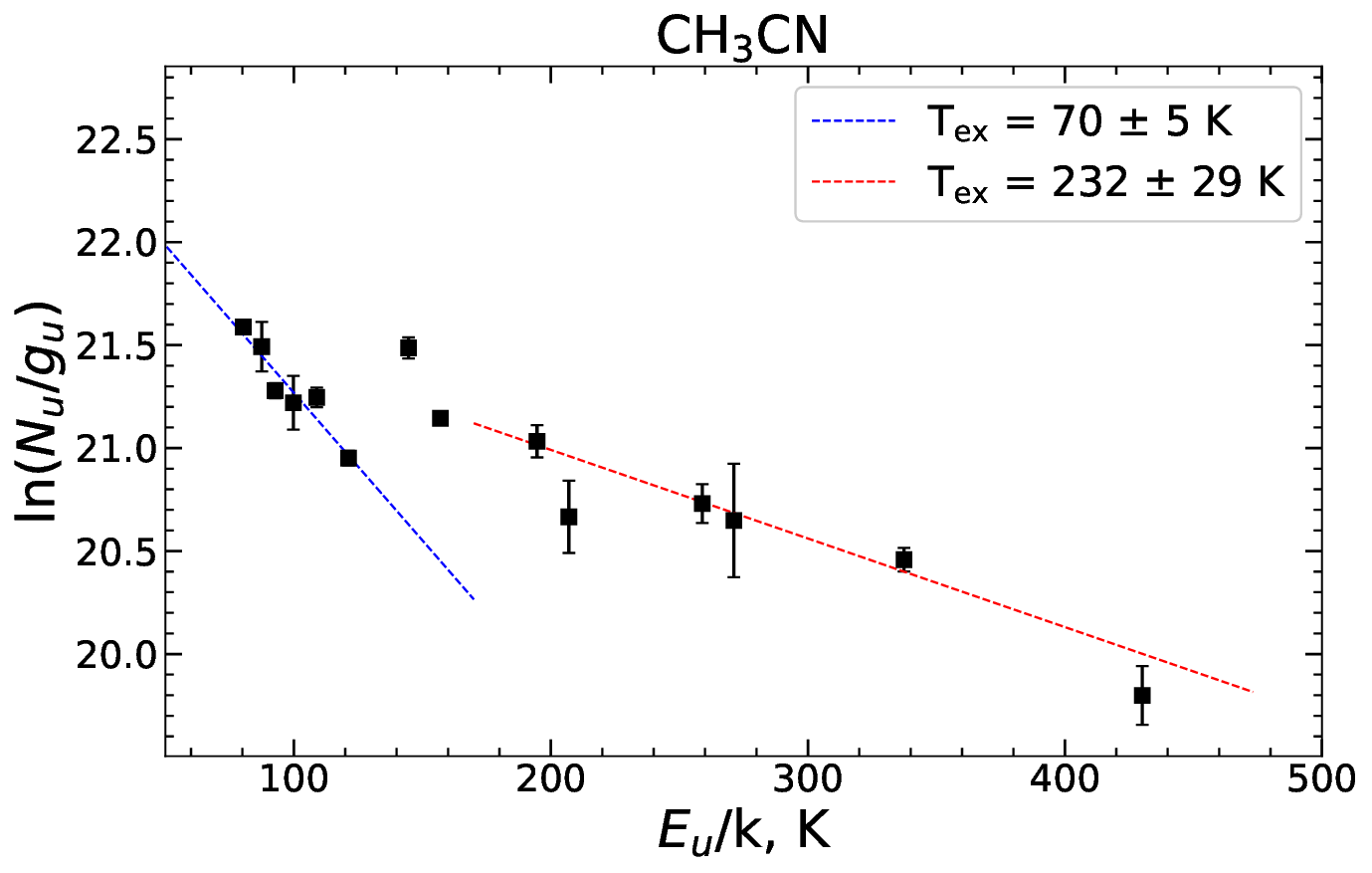}
     \includegraphics[width=0.96\linewidth]{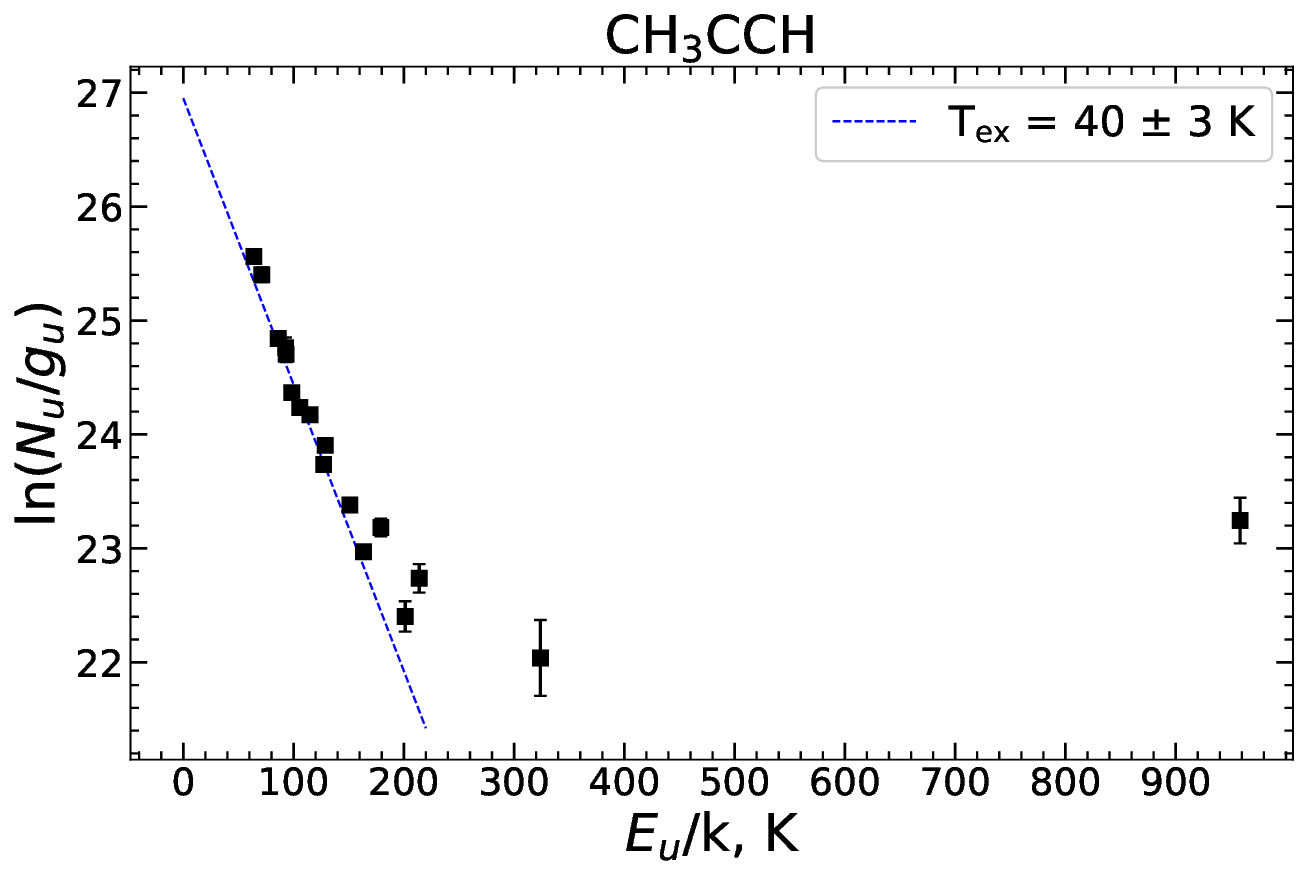}
    \caption{Rotation diagram for CH$_{3}$OH (top panel), CH$_{3}$CN (middle panel), and CH$_{3}$CCH (bottom panel) molecules for the RCW\,120~YSO~S2 source. The dashed line shows the approximation of the observational data by a linear function}
    \label{fig:rotation}
\end{figure}

\begin{figure*}[!ht]
    \centering
    \includegraphics[width=1.0\linewidth]{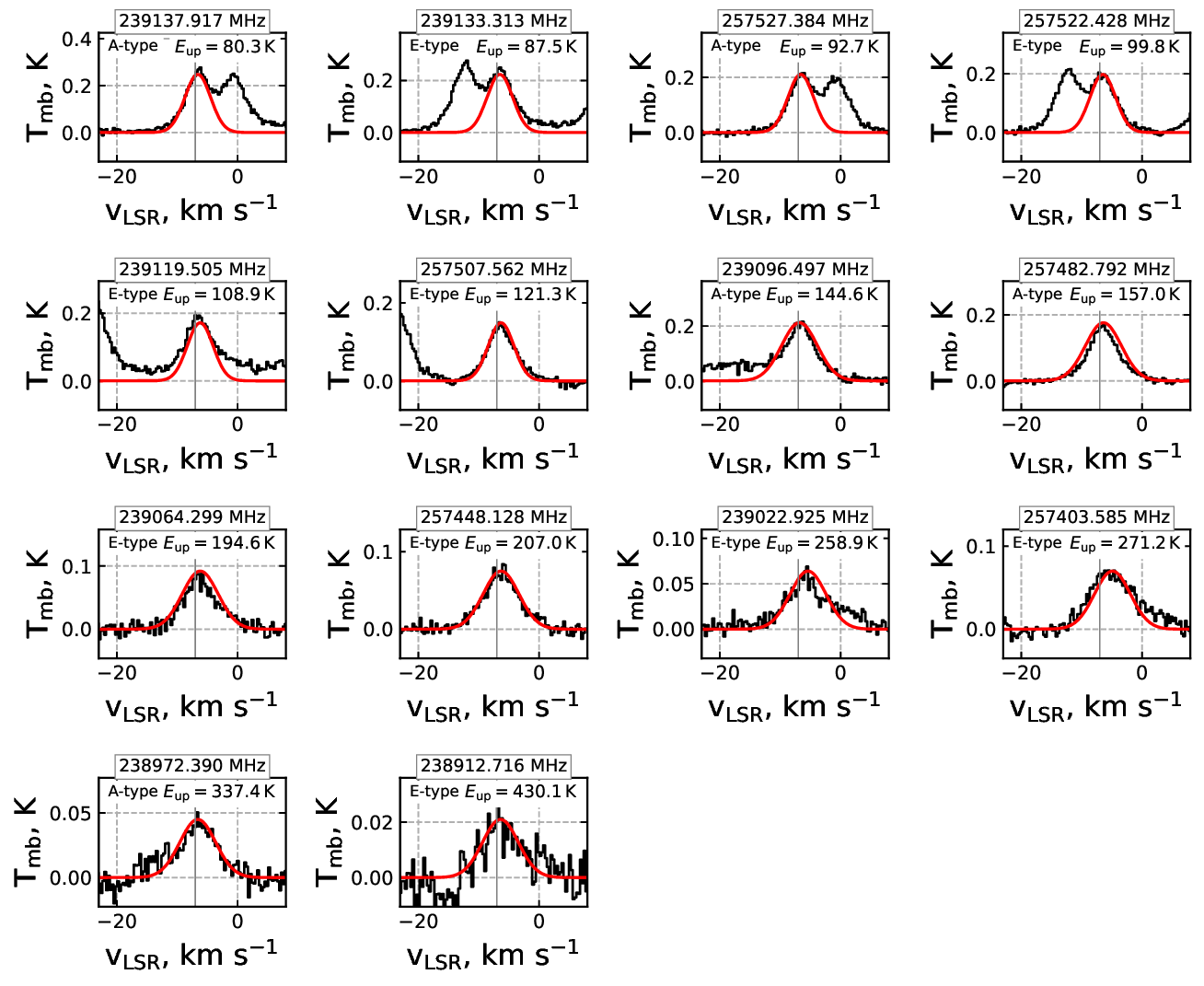}
    \caption{CH$_{3}$CN line profiles for RCW\,120~YSO~S2. The LSR velocity of -7 km s$^{-1}$ is indicated by a vertical solid gray line. When approximating using the Gaussian function (red), for lines with E$_{\rm up}$ $\leq$ 157 K, the line width was fixed at $\Delta$v=5 km s$^{-1}$. For lines with E$_{\rm up}$ > 157 K, $\Delta$v=7 km s$^{-1}$}
    \label{fig:profiles_ch3cn}
\end{figure*}

\begin{figure*}[!ht]
    \centering
    \includegraphics[width=1.0\linewidth]{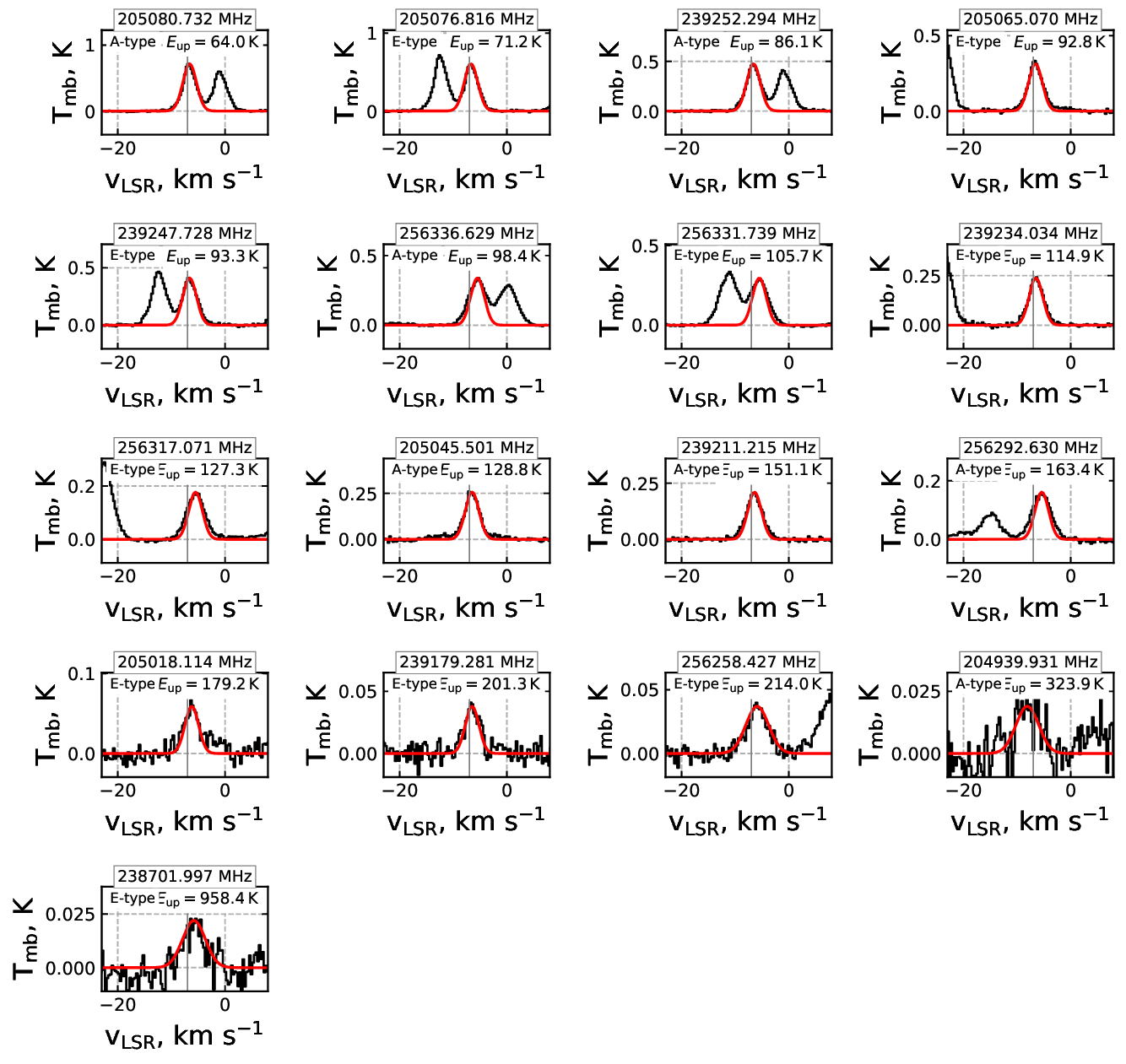}
    \caption{CH$_{3}$CCH line profiles for RCW\,120~YSO~S2. The LSR velocity of -7 km s$^{-1}$ is indicated by a vertical solid gray line. When approximating using the Gaussian function (red color) for lines with E$_{\rm up}$ $\leq$ 201.3 K, the line width was fixed at $\Delta$v=3 km s$^{-1}$. For lines with E$_{\rm up}$ > 201.3 K, $\Delta$v=5 km s$^{-1}$}
    \label{fig:profiles_ch3cch}
\end{figure*}

\begin{figure*}[!ht]
    \centering
    \includegraphics[width=1.0\linewidth]{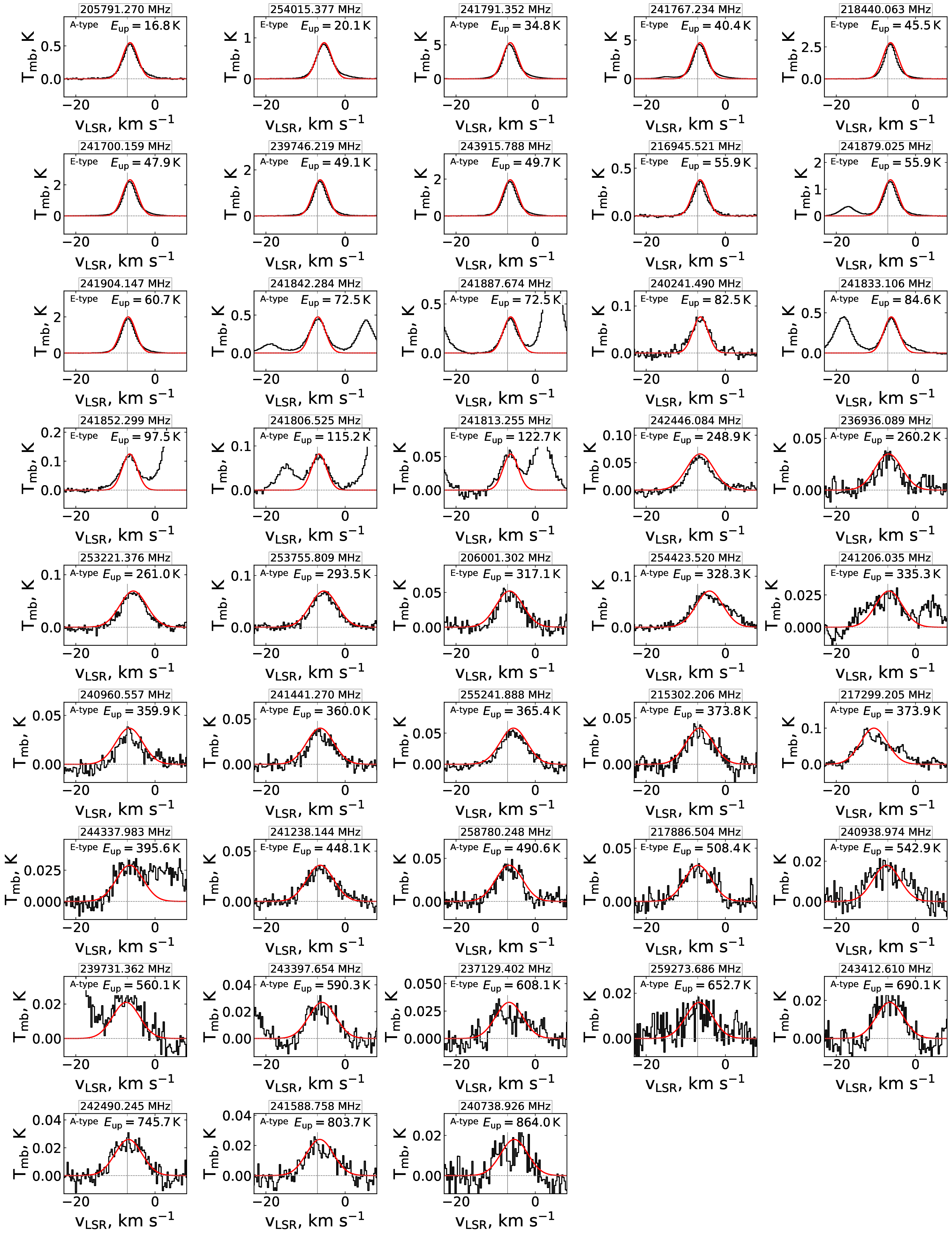}
    \caption{Profiles of CH$_{3}$OH lines for RCW\,120~YSO~S2. The LSR velocity of -7 km s$^{-1}$ is indicated by a vertical solid gray line. When approximating using the Gaussian function (red color) for lines with E$_{\rm up}$ $\leq$ 122.7 K, the line width was fixed at $\Delta$v=4.5 km s$^{-1}$. For lines with E$_{\rm up}$ > 122.7 K, $\Delta$v=8 km s$^{-1}$}
    \label{fig:profiles_ch3oh}
\end{figure*}

As noted above, the largest number of lines in the observed range belong to CH$_{3}$CN, CH$_{3}$CCH, and CH$_{3}$OH molecules. For this molecules we found differences in spectral profiles with dependence on the $E_{\rm u}$. Transitions with energies lower than the threshold values $E_{\rm u} \leq 157$~K for CH$_{3}$CN, $E_{\rm u} \leq 201.3$~K for CH$_{3}$CCH, $E_{\rm u} \leq 122.7$~K for CH$_{3}$OH, show a prominent central maximum. In contrast, transitions with $E_{\rm u}$ higher than these values are characterized by a noticeable broadening of the line profiles. For example, the profiles of the CH$_{3}$CN molecule lines with the Gaussian profile are shown in Fig.~\ref{fig:profiles_ch3cn}. We fit Gaussian function with fixed line width. For the lines with $E_{\rm u} \leq 157$ K, the line width was fixed at $\Delta v=5$ km s$^{-1}$, and for lines with $E_{\rm u} > 157$ K, $\Delta v =7$ km s$^{-1}$. The profiles of the CH$_{3}$OH molecule lines (see Fig.~\ref{fig:profiles_ch3oh}) show a similar separation, but the line widths and $E_{\rm u}$ value differ from those for CH$_{3}$CN. For line profiles with $E_{\rm u} \leq  122.7$ K, a value of $\Delta v = 4.5$ km s$^{-1}$ was used, while for transitions with higher energies, a value of $\Delta v = 8$ km s$^{-1}$ was used. The transitions of the CH$_{3}$CCH molecule were similarly divided according to the type of line profiles (see Fig.~\ref{fig:profiles_ch3cch}). Transitions with $E_{\rm u} \leq 201.3$ K were approximated with Gaussian profile with fixed $\Delta v = 3$ km s$^{-1}$, and for lines with $E_{\rm u}$ $>$ 201.3 K with $\Delta v = 5$~km s$^{-1}$. The fit parameters of CH$_{3}$OH, CH$_{3}$CN, CH$_{3}$CCH lines are given in the Table~\ref{tab:molecular_parameters_rotdi}.

\begin{table*}[!ht]
\caption{Estimates of excitation temperature and column density obtained for RCW\,120~YSO~S2 in the LTE approximation. Two values of excitation temperature and column density are given for the low-energy (l) and high-energy (h) components}\label{tab:rot_params}
\begin{center}
\begin{tabular}{ccccc}
\hline
\centering
\multirow{2}{*}{Molecule} & T$_{ex}$$^{\rm l}$, & T$_{ex}$$^{\rm h}$,&{N$^{\rm l}$ $\times$ 10$^{14}$,}& N$^{\rm h}$ $\times$ 10$^{14}$,\\
         &  K  & K & cm$^{-2}$  & cm$^{-2}$ \\
\hline

CH$_{3}$OH &27 $\pm$ 6 &238 $\pm$ 7 &13.20 $\pm$ 8.76  & 12.90 $\pm$ 7.23  \\

CH$_{3}$CN &70 $\pm$ 5 &232 $\pm$ 29 &0.11 $\pm$ 0.01  & 0.29 $\pm$ 0.06  \\

CH$_{3}$CCH & 40 $\pm$ 3 & & 2.96 $\pm$ 0.61 &  \\
\hline
\end{tabular}
\end{center}
\end{table*}

We used the method of rotational diagram analysis to determine the T$_{\rm ex}$ and N. Integral intensities of CH$_{3}$OH, CH$_{3}$CN, and CH$_{3}$CCH were calculated at the line widths mentioned above. Rotation diagrams for CH$_{3}$OH, CH$_{3}$CN, and CH$_{3}$CCH are shown in Fig.~\ref{fig:rotation}. The estimated T$_{\rm ex}$ and N are given in Table~\ref{tab:rot_params}. Since the diagrams show a visual separation of points into groups, each group was approximated separately. The obtained T$_{\rm ex}$ estimates show a division into two different components: hot and relatively cold. It indicates the presence of regions with different temperatures in the line of sight. For CH$_{3}$OH, the temperature separation is most prominent, since a larger number of high-energy transitions were identified for this molecule compared to the CH$_{3}$CN and CH$_{3}$CCH molecules. The rotational diagram for CH$_{3}$CN also shows a separation into two components.  The least prominent temperature separation is observed for the CH$_{3}$CCH molecule, whose rotational diagram shows only a low-energy component due to the unsufficient number (only 2) of lines with values $E_{\rm u} > 250$ K. 

\section{Discussion}
The studies \citet{2024AstBu..79..235P} and \citet{2025Plakitina} report the detection of 65 lines of 35 molecules in the direction of RCW\,120~YSO~S2 in a partly intersecting spectral range with a sensitivity of 40 mK. Manual identification was used in these works. It is almost not possible to manually identify lines in our spectra, whose sensitivity is an order of magnitude higher. In addition, manual search is time-consuming and does not take into account excitation conditions, therefore the choice of one molecule among all candidates may be ambiguous. In {\tt MOLLId}, a multi-level search allows molecules to be sequentially matched and transitions with high upper-level energy (E$_{\rm up} \geq 1000$~K) to be filtered out, which are presumably less likely for spectra of star-forming regions and molecular clouds.  

In some cases, the number of unidentified lines can be as high as $\sim$ 25\%, as, e.g., inthe work by~\citet{20111A&A...536A..33R}, or 20\% in~\citet{217A&A...603A.133A}. In some works, the number of unidentified lines is relatively small: 5\%~\citep{2006A&A...454..221B} and 1.8\%~\citep{2012A&A...546A..87Z}. All of the above works use {\tt XCLASS} as the main package for line identification. The efficiency of line identification is determined by the algorithms used in a package, as well as the availability and quality of spectroscopic data. At the same time, the kinematic structure of the observed objects can also affect the quality of identification.

In identification using {\tt MOLLId}, the number of unidentified lines for RCW\,120~YSO~S2 is 3 out of 407 lines (0.7\%), and for RCW\,120~YSO~S1 is 1 line out of 100. We suggest that the number of unidentified lines can be reduced by taking into account the vibrational states of different molecules or by changing the search ranges for the upper level energy or $\log A_{ij}$. Further testing of {\tt MOLLId} will help solve this problem.

Application of the LTE approximation can significantly improve the accuracy of the identification procedure. For example, in works by~\citet{gols} and \citet{2014ApJ...786...38H}, manual search is initially performed. After that the spectra are approximated using the {\tt XCLASS} tool. In the work by~\citet{Sewio_2018}, the {\tt MADCUBA}~\citep{martin} software package is used to automatically search for transitions and find the best match to the observed spectrum. Developement of the LTE approximation in {\tt MOLLId} will allow us to separate close transitions ($\Delta \nu \leq 0.0005$ GHz), overlapping lines, and more accurately select the most suitable transition for specific physical conditions. In further work, we plan to improve the accuracy of identification by adding an approximation of the observed spectra under the LTE conditions.

Thanks to the high sensitivity of the observations, the hypothesis that RCW\,120~YSO~S2 is on the hot core stage~\citep{2021MNRAS.503..633K} was confirmed. It is seen that the evaporation of dust grains begins due to the increase in dust temperature. Complex organic molecules start to appear in the gas phase. Meanwhile, in the works by \citet{2024AstBu..79..235P, 2025Plakitina}, no high-energy lines were detected and an analysis of the colder shell was performed. There relatively narrow lines were analyzed with widths of $4-5$~km s$^{-1}$. Our work convincingly shows the presence of hot gas in RCW\,120~YSO~S2. The fact that dust mantles are evaporating in this YSO can be confirmed by the presence of HDO lines in the spectrum (see Table \ref{tab:molecular_parameters_core2}). Our previous work \citep{2025arXiv250521945K} showed that evaporation began relatively recently.

The predominance of carbon-containing species and increased abundance of sulfur species is found in some YSOs. For example, the physical conditions and chemical composition of the hot cores were analyzed in the work by~\citet{2025arXiv250517403D}. The authors suggest that the close ($\sim$2$''$) sources MM1 and MM2 are at different evolutionary stages.  Since complex organic molecules are observed in one of the objects, while less complex species are observed in the other. According to our results, the sources RCW\,120~YSO~S1 and RCW\,120~YSO~S2 may also be at different evolutionary stages. Carbon-containing species were the most frequent in MM1 and MM2. Sulfur-containing species were the least frequent. Nitrogen-containing species were found to be twice as rare as oxygen-containing species. In our work, the number of identified sulfur-containing molecules is significantly higher than the number of nitrogen-containing molecules, which is contrary to the conclusions by the work~\citet{2025arXiv250517403D}, but qualitatively close to the increased number of sulfur-containing molecules compared to nitrogen-containing ones in SgrB2(N3) and SgrB2(N5) (see Fig.9 from~\citet{2025arXiv250517403D}).

Complex organic species such as CH$_{3}$CN, CH$_{3}$OH, and CH$_{3}$CCH are interpreted as indicators of the physical conditions of gas in the hot core. Due to their structural features, the rotational transitions of the CH$_{3}$CN molecule consist of closely spaced $K$-components. They trace the increase in the excitation energy of the transitions. CH$_{3}$CCH has a low excitation temperature compared to other complex organic molecules in the interstellar medium, such as CH$_{3}$CN and CH$_{3}$OCH$_{3}$. \citet{2019A&A...631A.137C} suggested that CH$_{3}$CCH is a molecule that traces the boundary between hot corinos and the cooler envelope and propagates further into the envelope. The abundance of CH$_{3}$CCH in various objects (see, e.g.,~\citet{Lin}) shows that this molecule is not sensitive to thermal desorption $\geq$ 100 K in hot core regions. In our work, CH$_{3}$CCH transitions were identified mostly with $E_{\rm u}< 300$ K. The $T_{\rm ex}$ value corresponds to the low-temperature ($<100$ K) component of the gas, which is consistent with the results of~\citet{Lin}.

The separation between low-energy and high-energy components for CH$_{3}$CN and CH$_{3}$OH molecules has been noted for some other objects and molecules. In the work by~\citet{Bouscasse} this separation is seen for CH$_{3}$CN, CH$_{3}$OH and many other complex organic molecules. \citet{Kalenskii_2023} found it for CH$_{3}$OH, and in~\citet{Wang} for the SO$_{2}$ molecule. The obtained separation is mainly interpreted as follows: low-energy lines appear in cold regions, while high-energy lines appear in hot regions. High-energy transitions, e.g., for CH$_{3}$OH, can be optically thick, which can affect the obtained splitting into two components and the estimates of the excitation temperature.

\section{Conclusion}
The study presents the {\tt MOLLId} software package, developed for the analysis of spectral data and the identification of molecular emission lines in the submillimeter and millimeter spectra. The software package is applied to six frequency ranges of observational spectra from 202 to 260 GHz of the RCW\,120~YSO~S2 and RCW\,120~YSO~S1 sources from the RCW 120 star-forming region. Based on the results of the study, we draw the following conclusions.

\begin{itemize}
    
    \item In the RCW\,120~YSO~S2 spectra, we identified 407 lines belonging to 79 molecules. Among the species we find small diatomic molecules (such as SO, C$^{34}$S, NS) and complex organic molecules with up to 9 atoms (e.g., CH$_{3}$OCH$_{3}$). The most frequent molecular line among all belongs to CH$_{3}$OH.
    
    \item The most numerous group of molecules are carbon-containing molecules. They are followed by oxygen-containing molecules, CH$_{3}$ groups, sulfur-containing molecules, and nitrogen-containing molecules. For the RCW\,120~YSO~S2 source the number of unidentified lines is 3 lines or 0.7\% of the total number of spectral lines at the level of $>$(5-7)$\sigma$. 
    
    \item In the spectra of the RCW\,120~YSO~S1 source, 100 lines of 40 molecules were identified: from small molecules (e.g., C$^{34}$S, SO, OCS, H$_{2}$S, H$_{2}$CS, SiO) to complex organic molecules (e.g., CH$_{3}$O$^{13}$CHO, CH$_{3}$COOH, CH$_{3}$CHO, CH$_{3}$OH, CH$_{3}$CN, CH$_{3}$CCH). For the the RCW\,120~YSO~S1 1 line or 1\% of the total number of lines remained unidentified.

    \item The detection of high-energy lines of the molecules CH$_{3}$OH, CH$_{3}$CN, and CH$_{3}$CCH confirmes the presence of a hot core in the RCW\,120~YSO~S2 region. These lines are not detected in the RCW\,120~YSO~S1 protostar, which indicates an earlier stage of star formation in it.
\end{itemize}

\section*{Fundings}

M. S. Kirsanova and A. A. Farafontova were supported by RSF, research project 24-22-00097.

\bibliographystyle{aa}
\bibliography{farafontova}

\begin{thebibliography}{35}
\expandafter\ifx\csname natexlab\endcsname\relax\def\natexlab#1{#1}\fi

\bibitem[{{Allen} {et~al.}(2017){Allen}, {van der Tak}, {S{\'a}nchez-Monge},
  {Cesaroni}, \& {Beltr{\'a}n}}]{217A&A...603A.133A}
{Allen}, V., {van der Tak}, F.~F.~S., {S{\'a}nchez-Monge}, {\'A}., {Cesaroni},
  R., \& {Beltr{\'a}n}, M.~T. 2017, \aap, 603, A133

\bibitem[{{Belitsky} {et~al.}(2018){Belitsky}, {Bylund}, {Desmaris}, {Ermakov},
  {Ferm}, {Fredrixon}, {Krause}, {Lapkin}, {Meledin}, {Pavolotsky}, {Rashid},
  {Shafiee}, {Strandberg}, {Sundin}, {Yadranjee Aghdam}, {Hesper}, {Barkhof},
  {Bekema}, {Adema}, {de Haan}, {Koops}, {Boland}, {Yagoubov}, {Marconi},
  {Siringo}, {Humphreys}, {Tan}, {Laing}, {Testi}, {Mroczkowski}, {Wild},
  {Saini}, \& {Bryerton}}]{2018AA...611A..98B}
{Belitsky}, V., {Bylund}, M., {Desmaris}, V., {et~al.} 2018, \aap, 611, A98

\bibitem[{{Beuther} {et~al.}(2006){Beuther}, {Zhang}, {Sridharan}, {Lee}, \&
  {Zapata}}]{2006A&A...454..221B}
{Beuther}, H., {Zhang}, Q., {Sridharan}, T.~K., {Lee}, C.~F., \& {Zapata},
  L.~A. 2006, \aap, 454, 221

\bibitem[{{Bouscasse, L.} {et~al.}(2024){Bouscasse, L.}, {Csengeri, T.},
  {Wyrowski, F.}, {Menten, K. M.}, \& {Bontemps, S.}}]{Bouscasse}
{Bouscasse, L.}, {Csengeri, T.}, {Wyrowski, F.}, {Menten, K. M.}, \& {Bontemps,
  S.} 2024, A\&A, 686, A252

\bibitem[{{Calcutt} {et~al.}(2019){Calcutt}, {Willis}, {J{\o}rgensen},
  {Bjerkeli}, {Ligterink}, {Coutens}, {M{\"u}ller}, {Garrod}, {Wampfler}, \&
  {Drozdovskaya}}]{2019A&A...631A.137C}
{Calcutt}, H., {Willis}, E.~R., {J{\o}rgensen}, J.~K., {et~al.} 2019, \aap,
  631, A137

\bibitem[{{Deharveng} {et~al.}(2009){Deharveng}, {Zavagno}, {Schuller},
  {Caplan}, {Pomar{\`e}s}, \& {De Breuck}}]{2009A&A...496..177D}
{Deharveng}, L., {Zavagno}, A., {Schuller}, F., {et~al.} 2009, \aap, 496, 177

\bibitem[{{Duan} {et~al.}(2025){Duan}, {Gou}, {Liu}, {Xu}, {Xu}, {Lan}, {Wang},
  {Pagani}, {Quan}, {Wang}, {Liu}, \& {He}}]{2025arXiv250517403D}
{Duan}, C., {Gou}, Q., {Liu}, T., {et~al.} 2025, arXiv e-prints,
  arXiv:2505.17403

\bibitem[{Erb(2024)}]{erb_2024}
Erb, D. 2024, pybaselines: A Python library of algorithms for the baseline
  correction of experimental data

\bibitem[{{Figueira, M.} {et~al.}(2017){Figueira, M.}, {Zavagno, A.},
  {Deharveng, L.}, {Russeil, D.}, {Anderson, L. D.}, {Men’shchikov, A.},
  {Schneider, N.}, {Hill, T.}, {Motte, F.}, {Mège, P.}, {LeLeu, G.}, {Roussel,
  H.}, {Bernard, J.-P.}, {Traficante, A.}, {Paradis, D.}, {Tigé, J.}, {André,
  P.}, {Bontemps, S.}, \& {Abergel, A.}}]{figueira}
{Figueira, M.}, {Zavagno, A.}, {Deharveng, L.}, {et~al.} 2017, Astronomy and
  Astrophysics, 600, A93

\bibitem[{{Gieser} {et~al.}(2019){Gieser}, {Semenov}, {Beuther}, {Ahmadi},
  {Mottram}, {Henning}, {Beltran}, {Maud}, {Bosco}, {Leurini}, {Peters},
  {Klaassen}, {Kuiper}, {Feng}, {Urquhart}, {Moscadelli}, {Csengeri},
  {Lumsden}, {Winters}, {Suri}, {Zhang}, {Pudritz}, {Palau}, {Menten},
  {Galvan-Madrid}, {Wyrowski}, {Schilke}, {S{\'a}nchez-Monge}, {Linz},
  {Johnston}, {Jim{\'e}nez-Serra}, {Longmore}, \&
  {M{\"o}ller}}]{2019A&A...631A.142G}
{Gieser}, C., {Semenov}, D., {Beuther}, H., {et~al.} 2019, \aap, 631, A142

\bibitem[{{Ginsburg} \& {Mirocha}(2011)}]{2011ascl.soft09001G}
{Ginsburg}, A. \& {Mirocha}, J. 2011, {PySpecKit: Python Spectroscopic
  Toolkit}, Astrophysics Source Code Library, record ascl:1109.001

\bibitem[{{Golshan, Roya Hamedani} {et~al.}(2024){Golshan, Roya Hamedani},
  {Sánchez-Monge, Álvaro}, {Schilke, Peter}, {Sewiło, Marta}, {Möller,
  Thomas}, {Veena, Vadamattom S.}, \& {Fuller, Gary A.}}]{gols}
{Golshan, Roya Hamedani}, {Sánchez-Monge, Álvaro}, {Schilke, Peter}, {et~al.}
  2024, A\&A, 688, A3

\bibitem[{{G{\"u}sten} {et~al.}(2006){G{\"u}sten}, {Nyman}, {Schilke},
  {Menten}, {Cesarsky}, \& {Booth}}]{2006AA...454L..13G}
{G{\"u}sten}, R., {Nyman}, L.~{\r{A}}., {Schilke}, P., {et~al.} 2006, \aap,
  454, L13

\bibitem[{{Haud}(2000)}]{2000AA...364...83H}
{Haud}, U. 2000, \aap, 364, 83

\bibitem[{{Henshaw} {et~al.}(2016){Henshaw}, {Longmore}, {Kruijssen}, {Davies},
  {Bally}, {Barnes}, {Battersby}, {Burton}, {Cunningham}, {Dale}, {Ginsburg},
  {Immer}, {Jones}, {Kendrew}, {Mills}, {Molinari}, {Moore}, {Ott}, {Pillai},
  {Rathborne}, {Schilke}, {Schmiedeke}, {Testi}, {Walker}, {Walsh}, \&
  {Zhang}}]{2016MNRAS.457.2675H}
{Henshaw}, J.~D., {Longmore}, S.~N., {Kruijssen}, J.~M.~D., {et~al.} 2016,
  \mnras, 457, 2675

\bibitem[{{Hern{\'a}ndez-Hern{\'a}ndez}
  {et~al.}(2014){Hern{\'a}ndez-Hern{\'a}ndez}, {Zapata}, {Kurtz}, \&
  {Garay}}]{2014ApJ...786...38H}
{Hern{\'a}ndez-Hern{\'a}ndez}, V., {Zapata}, L., {Kurtz}, S., \& {Garay}, G.
  2014, \apj, 786, 38

\bibitem[{{Juvela, Mika} \& {Tharakkal, Devika}(2024)}]{juvela}
{Juvela, Mika} \& {Tharakkal, Devika}. 2024, Astronomy and Astrophysics, 685,
  A164

\bibitem[{Kalenskii \& Mikheeva(2023)}]{Kalenskii_2023}
Kalenskii, S.~V. \& Mikheeva, E.~A. 2023, Astronomy Reports, 67, 1303–1328

\bibitem[{{Kirsanova} \& {Farafontova}(2025)}]{2025arXiv250521945K}
{Kirsanova}, M.~S. \& {Farafontova}, A.~A. 2025, Astronomy Letters, 51, 111

\bibitem[{{Kirsanova} {et~al.}(2021){Kirsanova}, {Salii}, {Kalenskii}, {Wiebe},
  {Sobolev}, \& {Boley}}]{2021MNRAS.503..633K}
{Kirsanova}, M.~S., {Salii}, S.~V., {Kalenskii}, S.~V., {et~al.} 2021, \mnras,
  503, 633

\bibitem[{{Lin, Y.} {et~al.}(2022){Lin, Y.}, {Wyrowski, F.}, {Liu, H. B.},
  {Izquierdo, A. F.}, {Csengeri, T.}, {Leurini, S.}, \& {Menten, K. M.}}]{Lin}
{Lin, Y.}, {Wyrowski, F.}, {Liu, H. B.}, {et~al.} 2022, A\&A, 658, A128

\bibitem[{Lindner {et~al.}(2015)Lindner, Vera-Ciro, Murray, Stanimirović,
  Babler, Heiles, Hennebelle, Goss, \& Dickey}]{Lindner_2015}
Lindner, R.~R., Vera-Ciro, C., Murray, C.~E., {et~al.} 2015, The Astronomical
  Journal, 149, 138

\bibitem[{{Martín, S.} {et~al.}(2019){Martín, S.}, {Martín-Pintado, J.},
  {Blanco-Sánchez, C.}, {Rivilla, V. M.}, {Rodríguez-Franco, A.}, \&
  {Rico-Villas, F.}}]{martin}
{Martín, S.}, {Martín-Pintado, J.}, {Blanco-Sánchez, C.}, {et~al.} 2019,
  Astronomy and Astrophysics, 631, A159

\bibitem[{{Möller, T.} {et~al.}(2017){Möller, T.}, {Endres, C.}, \& {Schilke,
  P.}}]{moller}
{Möller, T.}, {Endres, C.}, \& {Schilke, P.} 2017, Astronomy and Astrophysics,
  598, A7

\bibitem[{{Müller, H. S. P.} {et~al.}(2001){Müller, H. S. P.}, {Thorwirth,
  S.}, {Roth, D. A.}, \& {Winnewisser, G.}}]{cdms}
{Müller, H. S. P.}, {Thorwirth, S.}, {Roth, D. A.}, \& {Winnewisser, G.} 2001,
  Astronomy \& Astrophysics, 370, L49

\bibitem[{Pickett {et~al.}(1998)Pickett, Poynter, Cohen, Delitsky, Pearson, \&
  MÜLLER}]{PICKETT1998883}
Pickett, H., Poynter, R., Cohen, E., {et~al.} 1998, Journal of Quantitative
  Spectroscopy and Radiative Transfer, 60, 883

\bibitem[{{Plakitina} {et~al.}(2024){Plakitina}, {Kirsanova}, {Kalenskii},
  {Salii}, \& {Wiebe}}]{2024AstBu..79..235P}
{Plakitina}, K.~V., {Kirsanova}, M.~S., {Kalenskii}, S.~V., {Salii}, S.~V., \&
  {Wiebe}, D.~S. 2024, Astrophysical Bulletin, 79, 235

\bibitem[{{Plakitina} {et~al.}(2025){Plakitina}, {Kirsanova}, \&
  {Wiebe}}]{2025Plakitina}
{Plakitina}, K.~V., {Kirsanova}, M.~S., \& {Wiebe}, D.~S. 2025, Astrophysical
  Bulletin, 80, 348

\bibitem[{{Riener, M.} {et~al.}(2019){Riener, M.}, {Kainulainen, J.}, {Henshaw,
  J. D.}, {Orkisz, J. H.}, {Murray, C. E.}, \& {Beuther, H.}}]{Riener}
{Riener, M.}, {Kainulainen, J.}, {Henshaw, J. D.}, {et~al.} 2019, Astronomy and
  Astrophysics, 628, A78

\bibitem[{{Rolffs} {et~al.}(2011){Rolffs}, {Schilke}, {Zhang}, \&
  {Zapata}}]{20111A&A...536A..33R}
{Rolffs}, R., {Schilke}, P., {Zhang}, Q., \& {Zapata}, L. 2011, \aap, 536, A33

\bibitem[{Sewiło {et~al.}(2018)Sewiło, Indebetouw, Charnley, Zahorecz,
  Oliveira, van Loon, Ward, Chen, Wiseman, Fukui, Kawamura, Meixner, Onishi, \&
  Schilke}]{Sewio_2018}
Sewiło, M., Indebetouw, R., Charnley, S.~B., {et~al.} 2018, The Astrophysical
  Journal Letters, 853, L19

\bibitem[{{Vastel} {et~al.}(2015){Vastel}, {Bottinelli}, {Caux}, {Glorian}, \&
  {Boiziot}}]{2015sf2a.conf..313V}
{Vastel}, C., {Bottinelli}, S., {Caux}, E., {Glorian}, J.~M., \& {Boiziot}, M.
  2015, in SF2A-2015: Proceedings of the Annual meeting of the French Society
  of Astronomy and Astrophysics, ed. F.~{Martins}, S.~{Boissier}, V.~{Buat},
  L.~{Cambr{\'e}sy}, \& P.~{Petit}, 313--316

\bibitem[{{Wang, K.-S.} {et~al.}(2013){Wang, K.-S.}, {Bourke, T. L.},
  {Hogerheijde, M. R.}, {van der Tak, F. F. S.}, {Benz, A. O.}, {Megeath, S.
  T.}, \& {Wilson, T. L.}}]{Wang}
{Wang, K.-S.}, {Bourke, T. L.}, {Hogerheijde, M. R.}, {et~al.} 2013, A\&A, 558,
  A69

\bibitem[{{Zavagno} {et~al.}(2007){Zavagno}, {Pomar{\`e}s}, {Deharveng},
  {Hosokawa}, {Russeil}, \& {Caplan}}]{2007A&A...472..835Z}
{Zavagno}, A., {Pomar{\`e}s}, M., {Deharveng}, L., {et~al.} 2007, \aap, 472,
  835

\bibitem[{{Zernickel} {et~al.}(2012){Zernickel}, {Schilke}, {Schmiedeke},
  {Lis}, {Brogan}, {Ceccarelli}, {Comito}, {Emprechtinger}, {Hunter}, \&
  {M{\"o}ller}}]{2012A&A...546A..87Z}
{Zernickel}, A., {Schilke}, P., {Schmiedeke}, A., {et~al.} 2012, \aap, 546, A87

\end{thebibliography}

\clearpage
\onecolumn
\section*{Appendix}
\tiny
\begin{longtable}{cccccccccl}

\caption{Parameters of identified lines in RCW\,120~YSO~S1. Observational ranges \textbf{B1} - \textbf{B6}. The code indicates the presence of the molecule only in the JPL catalog (1), in both catalogs (2) or only in the CDMS catalog (3). The lines which were not succesfully matched with the algorithm are labeled as "UNMATCHED"}
\label{tab:molecular_parameters_core1}\\

\hline
Molecule & Transition & $\nu$ & $E_u$ & $\int T_{\rm mb}dV$ & $\Delta V$ & $T_{\rm mb}$ & $v_{\rm LSR}$ & Catalog & Code \\
& & MHz & K & K km s$^{-1}$ & km s$^{-1}$ & mK & km s$^{-1}$ & & \\
\hline
\endfirsthead

\multicolumn{10}{c}{\textbf{Continuation of Table \ref{tab:molecular_parameters_core1}}} \\
\hline
Molecule & Transition & $\nu$ & $E_u$ & $\int T_{\rm mb}dV$ & $\Delta V$ & $T_{\rm mb}$ & $v_{\rm LSR}$ & Catalog & Code \\
& & MHz & K & K km s$^{-1}$ & km s$^{-1}$ & mK& km s$^{-1}$ & & \\
\hline
\endhead

\hline
\multicolumn{10}{r}{\textit{Continued on the next page}} \\
\endfoot

\hline
\endlastfoot

\multicolumn{10}{c}{B1}\\
SO$_{2}$ & $12_{0,12}$ - $11_{1,11}$ & 203391.550 & 70.12 & 0.17$\pm$0.01 & 4.73$\pm$0.40 & 34$\pm$2 & -6.2 & JPL & 2 \\
$^{33}$SO & $5_{4,6}$ - $4_{3,5}$ & 203939.255 & 38.30 & 0.24$\pm$0.02 & 5.06$\pm$0.55 & 44$\pm$2 & -8.4 & CDMS & 3 \\
CH$_{3}$CCH & $12_{3}$ - $11_{3}$ & 205045.501 & 129.00 & 0.20$\pm$0.01 & 4.02$\pm$0.26 & 47$\pm$2 & -6.6 & JPL & 2 \\
CH$_{3}$CCH & $12_{2}$ - $11_{2}$ & 205065.070 & 92.88 & 0.21$\pm$0.01 & 3.14$\pm$0.20 & 62$\pm$3 & -6.6 & JPL & 2 \\
CH$_{3}$CCH & $12_{1}$ - $11_{1}$ & 205076.816 & 71.20 & 0.49$\pm$0.03 & 3.45$\pm$0.21 & 132$\pm$6 & -6.9 & JPL & 2 \\
CH$_{3}$CCH & $12_{0}$ - $11_{0}$ & 205080.732 & 63.98 & 0.50$\pm$0.02 & 3.04$\pm$0.14 & 154$\pm$6 & -6.7 & JPL & 2 \\
SO$^{17}$O & $23_{7,17}$ - $24_{6,18}$ & 205159.894 & 365.11 & 0.06$\pm$0.01 & 2.51$\pm$0.73 & 23$\pm$3 & -9.2 & CDMS & 3 \\
\hline

\end{longtable}

\onecolumn

\begin{longtable}{cccccccccc}

\caption{Parameters of identified lines in RCW 120 YSO~S2. Observational ranges \textbf{B1} - \textbf{B6}. The code indicates the presence of the molecule only in the JPL catalog (1), in both catalogs (2) or only in the CDMS catalog (3)}
\label{tab:molecular_parameters_core2}\\

\hline
Molecule & Transition & $\nu$& $E_u$ & $\int T_{\rm mb}dV$ & $\Delta V$ & $T_{\rm mb}$ & $v_{\rm LSR}$ & Catalog & Code \\
& & MHz & K & K km s$^{-1}$ & km s$^{-1}$ & mK& km s$^{-1}$ & & \\
\hline
\endfirsthead

\multicolumn{10}{c}{\textbf{Continuation of Table \ref{tab:molecular_parameters_core2}}} \\
\hline
Molecule & Transition & $\nu$& $E_u$ & $\int T_{\rm mb}dV$ & $\Delta V$ & $T_{\rm mb}$ & $v_{\rm LSR}$ & Catalog & Code \\
& & MHz & K & K km s$^{-1}$ & km s$^{-1}$ & mK& km s$^{-1}$ & & \\
\hline
\endhead

\hline
\multicolumn{10}{r}{\textit{Continued on next page}} \\
\endfoot

\hline
\endlastfoot

\multicolumn{10}{c}{B1}\\
CH$_{3}$CN, v8=1 & $11_{-1,3}$ - $10_{1,3}$ & 203161.011 & 577.51 & 0.16$\pm$0.02 & 5.45$\pm$1.08 & 28$\pm$2 & -5.5 & JPL & 2 \\
CH$_{3}$CHO & $27_{2,25}$ - $27_{2,26}$ E, $vt$=1 & 203241.771 & 570.50 & 0.05$\pm$0.01 & 2.39$\pm$0.46 & 18$\pm$3 & -7.4 & JPL & 1 \\
CH$_{3}$OCH$_{3}$ & $3$ - $2$ & 203374.155 & 18.12 & 0.06$\pm$0.01 & 2.66$\pm$0.37 & 21$\pm$3 & -6.9 & JPL & 2 \\
CH$_{3}$COOH & $31_{31,0}$ - $31_{30,1}$ E & 203383.931 & 534.83 & 0.32$\pm$0.03 & 6.37$\pm$0.86 & 47$\pm$3 & -6.9 & CDMS & 3 \\
SO$_{2}$ & $12_{0,12}$ - $11_{1,11}$ & 203391.550 & 70.12 & 0.78$\pm$0.02 & 4.73$\pm$0.16 & 155$\pm$3 & -6.4 & JPL & 2 \\
CH$_{3}$OCH$_{3}$ & $3$ - $2$ & 203410.100 & 18.12 & 0.20$\pm$0.02 & 5.11$\pm$0.56 & 37$\pm$2 & -6.9 & JPL & 2 \\
CH$_{3}$OCH$_{3}$ & $3$ - $2$ & 203418.718 & 18.12 & 0.18$\pm$0.02 & 4.58$\pm$0.53 & 37$\pm$3 & -6.9 & JPL & 2 \\
\end{longtable}


\onecolumn

\begin{longtable}{ccccccccl}

\caption{Parameters of CH$_{3}$OH, CH$_{3}$CN, CH$_{3}$CN lines identified in RCW 120 YSO~S2 which were used in rotation diagram analysis}
\label{tab:molecular_parameters_rotdi}\\

\hline
Molecule & Transition & $\nu$, MHz & E$_{\rm u}$ & $\int T_{\rm mb}dV$ & $\Delta V$ & $T_{\rm mb}$ & $v_{\rm LSR}$ & Catalogue \\
& & & K & K km s$^{-1}$ & km s$^{-1}$ & mK & km s$^{-1}$ \\ 
\hline
\endfirsthead

\multicolumn{9}{c}{\textbf{Continuation of Table \ref{tab:molecular_parameters_core2}}} \\
\hline
Molecule & Transition & $\nu$, MHz & $E_{\rm u}$ & $\int T_{\rm mb}dV$ & $\Delta V$ & $T_{\rm mb}$ & $v_{\rm LSR}$ & Catalogue \\
& & & K & K km s$^{-1}$ & km s$^{-1}$ & mK & km s$^{-1}$ \\
\hline
\endhead

\hline
\multicolumn{9}{r}{\textit{Continued on next page}} \\
\endfoot

\hline
\endlastfoot

\multicolumn{9}{c}{CH$_{3}$OH}\\
CH$_{3}$OH & 12$_{-5,7}$ --- 11$_{-4,9}$E & 206001.302 & 317.1 & 442.78$\pm$50.60 & 8.00 & 52.00$\pm$6.00 & -6.5 & JPL \\
CH$_{3}$OH & 5$_{0,5}$ --- 4$_{0,4}$A & 241791.352 & 34.8 & 25534.08$\pm$504.71 & 4.50 & 5331.00$\pm$77.00 & -6.3 & JPL \\
CH$_{3}$OH & 5$_{1,5}$ --- 4$_{1,4}$E & 241767.234 & 40.4 & 22372.85$\pm$465.02 & 4.50 & 4671.00$\pm$63.00 & -6.4 & JPL \\
CH$_{3}$OH & 4$_{-2,3}$ --- 3$_{-1,2}$E & 218440.063 & 45.5 & 13569.32$\pm$201.29 & 4.50 & 2833.00$\pm$23.00 & -6.3 & JPL \\
CH$_{3}$OH & 5$_{-0,5}$ --- 4$_{-0,4}$E & 241700.159 & 47.9 & 11160.08$\pm$151.10 & 4.50 & 2330.00$\pm$8.00 & -6.3 & JPL \\
CH$_{3}$OH & 5$_{1,5}$ --- 4$_{1,4}$A & 239746.219 & 49.1 & 7534.25$\pm$96.32 & 4.50 & 1573.00$\pm$4.00 & -6.3 & JPL \\
CH$_{3}$OH & 5$_{1,4}$ --- 4$_{1,3}$A & 243915.788 & 49.7 & 9411.83$\pm$127.76 & 4.50 & 1965.00$\pm$5.00 & -6.3 & JPL \\
CH$_{3}$OH & 5$_{-1,4}$ --- 4$_{-2,3}$E & 216945.521 & 55.9 & 1815.31$\pm$43.22 & 4.50 & 379.00$\pm$7.00 & -6.4 & JPL \\
CH$_{3}$OH & 5$_{-1,4}$ --- 4$_{-1,3}$E & 241879.025 & 55.9 & 6485.30$\pm$107.20 & 4.50 & 1354.00$\pm$3.00 & -6.3 & JPL \\
CH$_{3}$OH & 5$_{2,3}$ --- 4$_{2,2}$E & 241904.147 & 60.7 & 9598.63$\pm$105.02 & 4.50 & 2004.00$\pm$5.00 & -6.8 & JPL \\
CH$_{3}$OH & 5$_{2,4}$ --- 4$_{2,3}$A & 241842.284 & 72.5 & 2284.70$\pm$193.90 & 4.50 & 477.00$\pm$4.00 & -6.9 & JPL \\
CH$_{3}$OH & 5$_{2,3}$ --- 4$_{2,2}$A & 241887.674 & 72.5 & 1767.41$\pm$683.28 & 4.50 & 369.00$\pm$3.00 & -6.2 & JPL \\
CH$_{3}$OH & 5$_{-3,3}$ --- 4$_{-2,5}$E & 240241.490 & 82.5 & 368.81$\pm$16.39 & 4.50 & 77.00$\pm$3.00 & -6.3 & JPL \\
CH$_{3}$OH & 5$_{3,2}$ --- 4$_{3,1}$A & 241833.106 & 84.6 & 2155.38$\pm$69.81 & 4.50 & 450.00$\pm$4.00 & -6.1 & JPL \\
CH$_{3}$OH & 5$_{-3,3}$ --- 4$_{-3,2}$E & 241852.299 & 97.5 & 598.72$\pm$15.97 & 4.50 & 125.00$\pm$2.00 & -6.4 & JPL \\
CH$_{3}$OH & 5$_{4,1}$ --- 4$_{4,0}$A & 241806.525 & 115.2 & 392.76$\pm$16.36 & 4.50 & 82.00$\pm$3.00 & -6.6 & JPL \\
CH$_{3}$OH & 5$_{4,2}$ --- 4$_{4,1}$E & 241813.255 & 122.7 & 258.65$\pm$14.78 & 4.50 & 54.00$\pm$2.00 & -6.2 & JPL \\
CH$_{3}$OH & 14$_{1,14}$ --- 13$_{2,11}$E& 242446.084 & 248.9 & 562.00$\pm$24.43 & 8.00 & 66.00$\pm$2.00 & -6.4 & JPL \\
CH$_{3}$OH & 14$_{1,13}$ --- 13$_{2,12}$A & 236936.089 & 260.2 & 298.03$\pm$42.58 & 8.00 & 35.00$\pm$4.00 & -6.8 & JPL \\
CH$_{3}$OH & 13$_{3,11}$ --- 13$_{2,12}$A & 253221.376 & 261.0 & 587.54$\pm$246.77 & 8.00 & 69.00$\pm$29.00 & -5.5 & JPL \\
CH$_{3}$OH & 14$_{3,12}$ --- 14$_{2,13}$A & 253755.809 & 293.5 & 596.06$\pm$23.37 & 8.00 & 70.00$\pm$3.00 & -5.4 & JPL \\
CH$_{3}$OH & 15$_{3,13}$ --- 15$_{2,14}$A & 254423.520 & 328.3 & 604.57$\pm$32.68 & 8.00 & 71.00$\pm$3.00 & -4.0 & JPL \\
CH$_{3}$OH & 5$_{-0,5}$ --- 4$_{-0,4}$E & 241206.035 & 335.3 & 238.42$\pm$25.10 & 8.00 & 28.00$\pm$2.00 & -6.7 & JPL \\
CH$_{3}$OH & 5$_{1,5}$ --- 4$_{1,4}$A & 240960.557 & 359.9 & 315.06$\pm$27.40 & 8.00 & 37.00$\pm$3.00 & -6.5 & JPL \\
CH$_{3}$OH & 5$_{1,4}$ --- 4$_{1,3}$A & 241441.270 & 360.0 & 340.60$\pm$12.16 & 8.00 & 40.00$\pm$1.00 & -6.3 & JPL \\
CH$_{3}$OH & 16$_{3,14}$ --- 16$_{2,15}$A & 255241.888 & 365.4 & 493.87$\pm$94.07 & 8.00 & 58.00$\pm$11.00 & -5.5 & JPL \\
CH$_{3}$OH & 6$_{1,6}$ - 7$_{2,6}$A & 215302.206 & 373.8 & 332.09$\pm$33.21 & 8.00 & 39.00$\pm$4.00 & -6.5 & JPL \\
CH$_{3}$OH & 6$_{1,5}$ --- 7$_{2,5}$A & 217299.205 & 373.9 & 851.51$\pm$56.23 & 8.00 & 100.00$\pm$4.00 & -10.5 & JPL \\
CH$_{3}$OH & 9$_{-1,9}$ --- 8$_{-0,8}$E & 244337.983 & 395.6 & 246.94$\pm$59.86 & 8.00 & 29.00$\pm$5.00 & -6.4 & JPL \\
CH$_{3}$OH & 5$_{1,4}$ --- 4$_{1,3}$E & 241238.144 & 448.1 & 306.54$\pm$19.78 & 8.00 & 36.00$\pm$2.00 & -6.3 & JPL \\
CH$_{3}$OH & 19$_{3,17}$ --- 19$_{2,18}$A & 258780.248 & 490.6 & 357.63$\pm$46.15 & 8.00 & 42.00$\pm$5.00 & -6.6 & JPL \\
CH$_{3}$OH & 20$_{-1,19}$ --- 19$_{-0,20}$E & 217886.504 & 508.4 & 289.51$\pm$28.95 & 8.00 & 34.00$\pm$3.00 & -6.9 & JPL \\
CH$_{3}$OH & 5$_{0,5}$ --- 4$_{0,4}$A & 240938.974 & 542.9 & 153.27$\pm$27.87 & 8.00 & 18.00$\pm$3.00 & -7.3 & JPL \\
CH$_{3}$OH & 18$_{6,12}$ - 19$_{5,15}$A & 243397.654 & 590.3 & 229.91$\pm$31.35 & 8.00 & 27.00$\pm$3.00 & -5.8 & JPL \\
CH$_{3}$OH & 22$_{-1,21}$ --- 21$_{-0,22}$E & 237129.402 & 608.1 & 281.00$\pm$35.12 & 8.00 & 33.00$\pm$3.00 & -6.6 & JPL \\
CH$_{3}$OH & 17$_{2,15}$ --- 16$_{1,15}$A & 259273.686 & 652.7 & 136.24$\pm$41.92 & 8.00 & 16.00$\pm$4.00 & -6.6 & JPL \\
CH$_{3}$OH & 23$_{3,20}$ --- 23$_{2,21}$A & 243412.610 & 690.1 & 161.79$\pm$40.45 & 8.00 & 19.00$\pm$4.00 & -6.5 & JPL \\
CH$_{3}$OH & 24$_{3,21}$ --- 24$_{2,22}$A & 242490.245 & 745.7 & 221.39$\pm$19.25 & 8.00 & 26.00$\pm$2.00 & -6.7 & JPL \\
CH$_{3}$OH & 25$_{3,22}$ --- 25$_{2,23}$A & 241588.758 & 803.7 & 204.36$\pm$20.44 & 8.00 & 24.00$\pm$2.00 & -6.4 & JPL \\
CH$_{3}$OH & 26$_{3,23}$ --- 26$_{2,24}$A & 240738.926 & 864.0 & 153.27$\pm$32.84 & 8.00 & 18.00$\pm$3.00 & -5.3 & JPL \\

\multicolumn{9}{c}{CH$_{3}$CN}\\
CH$_{3}$CN & 13$_0$ - 12$_0$A & 239137.917 & 80.3 & 1319.84$\pm$43.27 & 5.00 & 248.00$\pm$4.00 & -6.6 & CDMS \\
CH$_{3}$CN & 13$_1$ - 12$_1$E & 239133.313 & 87.5 & 1192.11$\pm$143.05 & 5.00 & 224.00$\pm$13.00 & -6.5 & CDMS\\
CH$_{3}$CN & 14$_0$ - 13$_0$A & 257527.384 & 92.7 & 1122.93$\pm$40.83 & 5.00 & 211.00$\pm$4.00 & -6.6 & CDMS\\
CH$_{3}$CN & 14$_1$ - 13$_1$E & 257522.428 & 99.8 & 1053.74$\pm$137.44 & 5.00 & 198.00$\pm$11.00 & -6.5 & CDMS\\
CH$_{3}$CN & 13$_2$ - 12$_2$E & 239119.505 & 108.9 & 915.37$\pm$43.59 & 5.00 & 172.00$\pm$5.00 & -6.2 & CDMS\\
CH$_{3}$CN & 14$_2$ - 13$_2$E & 257507.562 & 121.3 & 792.97$\pm$27.66 & 5.00 & 149.00$\pm$5.00 & -6.3 & CDMS\\
CH$_{3}$CN & 13$_3$ - 12$_3$A & 239096.497 & 144.6 & 1128.25$\pm$57.65 & 5.00 & 212.00$\pm$9.00 & -6.8 & CDMS\\
CH$_{3}$CN & 14$_3$ - 13$_3$A & 257482.792 & 157.0 & 936.66$\pm$26.26 & 5.00 & 176.00$\pm$4.00 & -6.4 & CDMS\\
CH$_{3}$CN & 13$_4$ - 12$_4$E & 239064.299 & 194.6 & 685.46$\pm$53.55 & 7.00 & 92.00$\pm$7.00 & -6.2 & CDMS\\
CH$_{3}$CN & 14$_4$ - 13$_4$E & 257448.128 & 207.0 & 558.80$\pm$98.04 & 7.00 & 75.00$\pm$13.00 & -6.3 & CDMS\\
CH$_{3}$CN & 13$_5$ - 12$_5$E & 239022.925 & 258.9 & 476.84$\pm$44.99 & 7.00 & 64.00$\pm$5.00 & -5.4 & CDMS\\
CH$_{3}$CN & 14$_5$ - 13$_5$E & 257403.585 & 271.2 & 521.55$\pm$143.61 & 7.00 & 70.00$\pm$19.00 & -4.9 & CDMS\\
CH$_{3}$CN & 13$_6$ - 12$_6$A & 238972.390 & 337.4 & 335.28$\pm$19.16 & 7.00 & 45.00$\pm$2.00 & -6.6 & CDMS\\
CH$_{3}$CN & 13$_7$ - 12$_7$E & 238912.716 & 430.1 & 156.46$\pm$22.35 & 7.00 & 21.00$\pm$3.00 & -6.4 & CDMS\\
\multicolumn{9}{c}{CH$_{3}$CCH}\\
CH$_{3}$CCH & 12$_0$ - 11$_0$A & 205080.732 & 64.0 & 2286.30$\pm$40.11 & 3.00 & 716.00$\pm$9.00 & -6.6 & CDMS\\
CH$_{3}$CCH & 12$_1$ - 11$_1$E & 205076.816 & 71.2 & 1935.05$\pm$125.78 & 3.00 & 606.00$\pm$26.00 & -6.7 & CDMS\\
CH$_{3}$CCH & 14$_0$ - 13$_0$A & 239252.294 & 86.1 & 1516.75$\pm$27.41 & 3.00 & 475.00$\pm$5.00 & -6.6 & CDMS\\
CH$_{3}$CCH & 12$_2$ - 11$_2$E & 205065.070 & 92.8 & 999.46$\pm$89.24 & 3.00 & 313.00$\pm$31.00 & -6.6 & CDMS\\
CH$_{3}$CCH & 14$_1$ - 13$_1$E & 239247.728 & 93.3 & 1312.39$\pm$88.08 & 3.00 & 411.00$\pm$17.00 & -6.6 & CDMS\\
CH$_{3}$CCH & 15$_0$ - 14$_0$A & 256336.629 & 98.4 & 1082.48$\pm$15.03 & 3.00 & 339.00$\pm$2.00 & -5.5 & CDMS\\
CH$_{3}$CCH & 15$_1$ - 14$_1$E & 256331.739 & 105.7 & 945.17$\pm$7.56 & 3.00 & 296.00$\pm$7.00 & -5.5 & CDMS\\
CH$_{3}$CCH & 14$_2$ - 13$_2$E & 239234.034 & 114.9 & 759.97$\pm$35.35 & 3.00 & 238.00$\pm$11.00 & -6.5 & CDMS\\
CH$_{3}$CCH & 15$_2$ - 14$_2$E & 256317.071 & 127.3 & 565.19$\pm$15.28 & 3.00 & 177.00$\pm$3.00 & -5.5 & CDMS\\
CH$_{3}$CCH & 12$_3$ - 11$_3$A & 205045.501 & 128.8 & 817.45$\pm$44.43 & 3.00 & 256.00$\pm$14.00 & -6.5 & CDMS\\
CH$_{3}$CCH & 14$_3$ - 13$_3$A & 239211.215 & 151.1 & 670.56$\pm$25.15 & 3.00 & 210.00$\pm$6.00 & -6.4 & CDMS\\
CH$_{3}$CCH & 15$_3$ - 14$_3$A & 256292.630 & 163.4 & 514.10$\pm$28.56 & 3.00 & 161.00$\pm$4.00 & -5.4 & CDMS\\
CH$_{3}$CCH & 12$_4$ - 11$_4$E & 205018.114 & 179.2 & 188.40$\pm$14.49 & 3.00 & 59.00$\pm$3.00 & -6.2 & CDMS\\
CH$_{3}$CCH & 14$_4$ - 13$_4$E & 239179.281 & 201.3 & 121.34$\pm$16.18 & 3.00 & 38.00$\pm$3.00 & -6.5 & CDMS\\
CH$_{3}$CCH & 15$_4$ - 14$_4$E & 256258.427 & 214.0 & 196.91$\pm$24.61 & 5.00 & 37.00$\pm$1.00 & -5.9 & CDMS\\
CH$_{3}$CCH & 12$_6$ - 11$_6$A & 204939.931 & 323.9 & 101.12$\pm$33.71 & 5.00 & 19.00$\pm$5.00 & -8.1 & CDMS\\
CH$_{3}$CCH & 14$_{11}$ - 13$_{11}$E & 238701.997 & 958.4 & 117.08$\pm$23.42 & 5.00 & 22.00$\pm$3.00 & -5.8 & CDMS\\

\end{longtable}

\end{document}